\documentclass[preprint,review,1p,times,10.5pt]{elsarticle}

\usepackage[bookmarks,bookmarksnumbered]{hyperref}
\hypersetup{colorlinks = true,linkcolor = blue,anchorcolor =red,citecolor = blue,filecolor = red,urlcolor = red, pdfauthor=author}
\usepackage{graphicx}
\usepackage{subfigure}
\usepackage{epstopdf}
\usepackage{caption}
\graphicspath{{./Figure/}}
\usepackage{enumitem}
\usepackage{float}
\usepackage{txfonts}
\usepackage{booktabs}
\usepackage{natbib}
\usepackage{siunitx}
\usepackage{xcolor}
\usepackage{bm}

\usepackage{multirow}

\usepackage[version=4]{mhchem}
\usepackage{ulem}

\usepackage{url}

\usepackage[T1]{fontenc}

\usepackage{amssymb}
\usepackage{amsmath}
\usepackage[capitalise]{cleveref}
\usepackage{array}

\usepackage{etoolbox} 

\journal{Journal of Energy Storage}

\begin{document}


\captionsetup[figure]{labelfont={bf}, labelformat={default}, labelsep=period, name={Fig.}}


\begin{frontmatter}



\title{Evaluation of tortuosity: A radical tessellation-based method in porous spherical particle packing systems}


\author[Tongji]{Zongli Chen}
\author[Tongji]{Chenzhe Li}
\author[Tongji]{Ying Zhao\corref{cor1}}
\cortext[cor1]{Corresponding author.}
\ead{19531@tongji.edu.cn}
\address[Tongji]{School of Aerospace Engineering and Applied Mechanics, Tongji University, Shanghai 200092, China}

\begin{abstract}
Estimating the tortuosity of porous electrodes is important for understanding the performance of lithium-ion batteries and optimizing the design of electrode microstructures. In this work, a new method for estimating the tortuosity of porous electrodes is proposed based on radical tessellation, and the results agree well with those calculated by empirical formulas and finite element simulations. Compared with empirical formulas, the proposed method can estimate tortuosity for more complicated microstructures, regardless of particle distribution and size dispersity; compared with finite element simulations, the proposed method  offers a significant advantage in computational efficiency. Based on the method, the influence of different microstructure characters on the tortuosity is discussed. It is found that in addition to the porosity, the particle size and particle aggregation morphology are all critical in influencing the tortuosity of the porous electrode. Finally, the tortuosity obtained by this method is integrated into the Pseudo-2-dimensional (P2D) model for the calculation of effective properties, and the results show that this method offers an efficient way in improving the prediction accuracy of P2D models.
\end{abstract}




\end{frontmatter}


\section{Introduction}\label{sec:intro}

\begin{table}[h]
\centering
\begin{tabular}{|l l l l|}
\hline
Nomenclature & Description & Nomenclature & Description \\ 
\hline
P2D           & Pseudo-2-dimension  & $\tau$           & tortuosity              \\ 
$\phi$           & porosity     & $p$           & characterized experimentally             \\ 
KC           & Kozeny-Carman            & LBM           & lattice Boltzmann method             \\ 
TBM           & tessellation-based method      &$L_{pi}$           & tortuous path of length             \\
$L$           & thickness of the domain        & $s_{i}$           & start node             \\
$T_{i}$           & target node        & $\bar\tau$           & mean tortuosity              \\
$V_{aggregation}$           & total volume of particles involved    &DBSCAN & Density-Based Spatial Clustering             \\
              &  in aggregation                       &               & of Applications with Noise       \\
$N$           & the total number of vertex pairs    & $V_{total}$           & total volume of the particles             \\
$R$           & radius of the real particle &  $V$           & electric voltage           \\
$D_{b}$       & the distance of each particle of interest      & $R_\varepsilon$    & the minimum distances between           \\
    & from the boundaries      &     & two particles             \\
$r$           & radius of the background particle     &  $\sigma_l^\mathrm{eff}$           & the effective conductivity             \\
$D_l^\mathrm{eff}$           & the effective diffusivity      & $\sigma_l$           & conductivity             \\
$D_l$           & Li ion diffusivity of the pore phase        &  $\varepsilon_l$           & Li ion diffusivity of the pore phase             \\
\hline
\end{tabular}
\end{table}

Most natural and artificial porous structures are chaotic at meso- or micro-scales, and the key to evaluate the performance of a porous structure is to effectively and correctly characterize its complexity. In many engineering and research areas, including oil and gas production, electrochemistry, geosciences, civil engineering, there is an urgent need to quantitatively evaluate the relationship between microstructure and its transport performance~\citep{Eloisa_2022_Energy122151,Moldrup_2001_SSAJ653613x,Yavuz_2023_MS02166,CAI_2019_Energy116051}. Quantities such as porosity, specific surface area, correlation function, pore size, connectivity and tortuosity are often used to characterise porous structures~\citep{FU_2021_ESR103439,LI_2023_Energy126456,JI_2023_Energy128628,Malik_2024_JES109937}. Among them, tortuosity, as one of the key parameters to represent the complex porous microstructure, has an important impact on the study of macroscopic transport properties of flow, diffusion, and infiltration in porous media~\citep{Saomoto_2015_TPM0467,Xu_2024_JES109939,LI_2024_PT119120}.

Lithium-ion batteries, fuel cells and supercapacitor, as typical electrochemical devices, are currently major power suppliers for electric vehicles, and their electrochemical performance is highly dependent on the microstructure of their porous electrodes~\citep{WU_2022_ACSNano2c00129,Lu_2018_EES1064A,HE_2020_JPS228771,Kadam_2023_JEM10019,Spray_CS202202504}. In particular, tortuosity is often used to correlate the effective ionic/electronic transport properties with microstructures, and to design a microstructure with low tortuosity is key to enhance the performance of electrochemical devices~\citep{Tjaden_2016_IMR1249995,CHEN_2024_JPS234095,Xie_2020_JES101837}. On the one hand, optimizing the microstructure of the electrode can effectively reduce the tortuosity, thereby decreasing electrode polarization while maintaining the capacity of electrodes. On the other hand, in numerical practices, such as in Pseudo-2-dimension (P2D)~\citep{Fuller_1994_JES2054684} and Adler-Lane-Steele models~\citep{Adler_1996_JES1837252}, tortuosity is often set as an input parameter for calculations, and it is shown that a correct estimation of tortuosity is important to obtain reliable simulation results. In most current models, the tortuosity is estimated as only depending on porosity~\citep{Ghanbarian_2013_SSSAJ0435}, as summarized in \cref{tb:empirical_models}, with $\tau$ and $\phi$ being the tortuosity and porosity, respectively. These models offer straightforward estimations of the tortuosity based on porosity, helping to gain a deeper understanding of the characteristics of porous media.
\newcounter{mycounter}
\setcounter{mycounter}{1}\label{model:1}
\begin{table}[h]
\centering
\caption{Representative empirical models for tortuosity estimation in circular/spherical particle packing structures}\label{tb:empirical_models}
\begin{tabular}{l c c c}
\hline\hline
Expression&Eq.&Description&Ref.\\\hline
$\tau= 1 - p\ln\phi$ & (\themycounter) & 2D or 3D structure, $p$ characterized experimentally (Comiti's model) & \citep{Ghanbarian_2013_SSSAJ0435,Weissberg_1963_JAP1729783,COMITI_1989_CES80031}\\
$\tau  = 1 + \frac{1}{2}\left(1 - \phi\right)$& \refstepcounter{mycounter}\label{model:2}(\themycounter) & 3D structure, based on Laplace Maxwell equation (Maxwell's model)&\citep{Ghanbarian_2013_SSSAJ0435}\\
$\tau  = {\phi ^{ - 0.5}}$& \refstepcounter{mycounter}\label{model:3}(\themycounter) & Bruggeman relation (Bruggeman's model)&\citep{Chung_2013_MSMSE074009,TJADEN_2016_COCE006}\\\hline\hline
\end{tabular}
\end{table}
\setcounter{equation}{\themycounter}

Currently, most studies on tortuosity in porous media focus on macro-scale porosity~\citep{Ghanbarian_2013_SSSAJ0089}. However, in addition to porosity, the microstructural heterogeneity in porous media also influences tortuosity, and disregarding heterogeneity can result in inaccurate estimations of tortuosity~\citep{ZOU_2023_Energy129512,Joseph_2022_SR23643,LI_2024_PT120129}. The effect of microstructural heterogeneity on tortuosity is not only reflected in pore shape and distribution but also involves differences in the nature of substances in different regions, such as local aggregation of particles. For example, in battery electrodes, the aggregation of additives and active particles leads to changes in local physical resistance, which, in turn, significantly affects the tortuosity of the migration paths of electrons and ions. These microscopic changes not only directly affect the overall tortuosity of the electrodes but also impact the interfacial impedance and power performance of the battery~\citep{Usseglio_2020_JES913b,Zhu_2011_JES3625286}. Therefore, it is crucial to propose a tortuosity estimation method that comprehensively considers these microscopic heterogeneous factors. By introducing multiscale modeling techniques and high-resolution microstructural analysis, the tortuosity variation of porous media under different microstructural conditions can be evaluated more accurately, providing a theoretical basis for the design and performance optimization of porous media materials~\citep{Selly_2013_Fra00138,Sun_2024_JES112940}.

As early efforts, Carman introduced the concept of tortuosity to explain the tortuous characteristics of flow through a particle bed, and combined the Kozeny-Carman (KC) equation and experiments to obtain the tortuosity~\citep{CARMAN_1997_CERD80003,WU_2023_Energy127723,LI_2022_IJES103658,CAO_2024_P535}. They measured porosity, permeability, and specific surface area experimentally and inverted the KC equation for the calculation of tortuosity. However, this value of tortuosity does not match the true tortuosity observed or measured under actual conditions, which can not be physically justified~\citep{Rehman_2024_ACME10094,RUAN_2022_JRMGE08010,Srisutthiyakorn_2017_Int0080}. Later, theoretical and numerical approaches have also been proposed, such as lattice percolation theory and lattice Boltzmann method (LBM)~\citep{YAN_2021_Energy120773,Ghanbarian_2013_SSSAJ0089,Vasseur_2021_PhysRevE062613}. These methods require a significant amount of computational resources to simulate the process~\citep{Graczyk_2020_SR78415}. 3D imaging techniques such as X-ray tomography and scanning electron microscopy tomography are effective methods for obtaining tortuosity. However, these methods entail significant computational expenses~\citep{Kehrwald_2011_JES079112,HUTZENLAUB_201377_EC006,Martin_2014_AEM01278,SHANTI_2014_ActaM003,ERCELIK_2023_Energy125531}, and the difficulties and costs to obtain the microstructure climb exponentially when the size of microstructures decreases to micro- or nano-scale~\citep{Ebner_2015_JES0111502}.

In order to alleviate calculation expenses, we would like to develop a method to estimate tortuosity that is basically from the topology of the structure, without involving physical processes. As we understand, the tortuosity is related to the length and curvature of the flow lines in porous media and is extremely sensitive to the distribution of obstacles~\citep{Berg_2014_TPM0307}. And the topological division of porous structures can effectively identify the positions and distributions of pores in the porous media. Therefore, it is possible to obtain the flow lines and curvature in porous media by means of topological division of the porous structure. Among them, Voronoi tessellation is a commonly employed method for topological division of porous structures, and it has been widely used in the topological analysis of spherical particle packing systems~\citep{Rycroft_2009_Chaos3215722,Yang_2002_PhysRevE041302}. The Voronoi tessellation can effectively characterize the pore network of particle packings systems and plays an important role in modeling porous structures with transport or transmission properties~\citep{Cheng_2013_IECR3033137,DONG_2016_CES013}. However, Voronoi tessellation is only applicable to single-sized particles packing systems, and cannot be used for particle packing systems with varied particle sizes~\citep{YI_2012_PT042}. For the multi-sized particle packing systems, the porous packing structure can be topological partitioned by radical tessellation~\citep{Yi_2015_PhysRevE032201}. Radical tessellation, as an extension of Voronoi tessellation, takes the size difference of the particles of a porous packing structure into account and partitions the porous particle packing structure topologically by constructing an isotropic plane of two neighboring particles~\citep{CHEN_2022_PT118002}. Overall, employing Radical tessellation for the topological division of porous packing structures not only enables the effective acquisition of length and curvature information of streamlines within porous media, but also efficiently captures effects induced by particle variations for porous packing systems with different sizes and particle distributions (obstacle distributions).

In this work, based on our previous work, a modified tessellation-based method (TBM) for evaluating the tortuosity of porous structures is developed, where radical tessellation is used for the topological division of the porous structures~\citep{CHEN_2024_JPS234095,CHEN_2022_PT118002}. The porous structures are assumed to be composed of circular (2D) or spherical (3D) particles, which enables the convenient acquisition of connected pore pathways through the structure. The target porous structure is obtained in the form of the particle packing, where the method of generation the particle packing structure is derived from our previous work~\citep{CHEN_2022_PT118002}. The porous structure is topologically partitioned by radical tessellation to obtain all paths in the porous structure that an object can flow or diffuse through. Further, the Dijkstra's search algorithm is used to find the shortest path through the porous structure and to calculate the tortuosity of the porous structure. In order to enhance the accuracy of tortuosity estimation with loosely packed porous structure, a group of pseudo background particles is introduced. The results show that adding background particles can effectively help to accurately estimate tortuosity. As a validation, the results from the TBM are compared with those obtained from finite element simulations and empirical formulas, which shows great agreement.  The proposed method is then employed for the tortuosity estimation of different heterogeneous porous microstructures, and it is found that the tortuosity is related to, in addition to porosity, the size of the particles composing the porous structure and the aggregation morphology of the particles. Finally, different forms of P2D parameter correction models are compared to verify the effectiveness of TBM in improving the prediction accuracy of P2D models. In summary, this work proposed a tessellation-based method for spherical particle packing systems, which is fast and reliable, and is able to be easily employed in 3D configurations.

\section{Methods: tortuosity estimation in 2D and 3D}\label{sec:Analytical method}
\begin{figure}[h]
   \centering
   \includegraphics[scale=0.35]{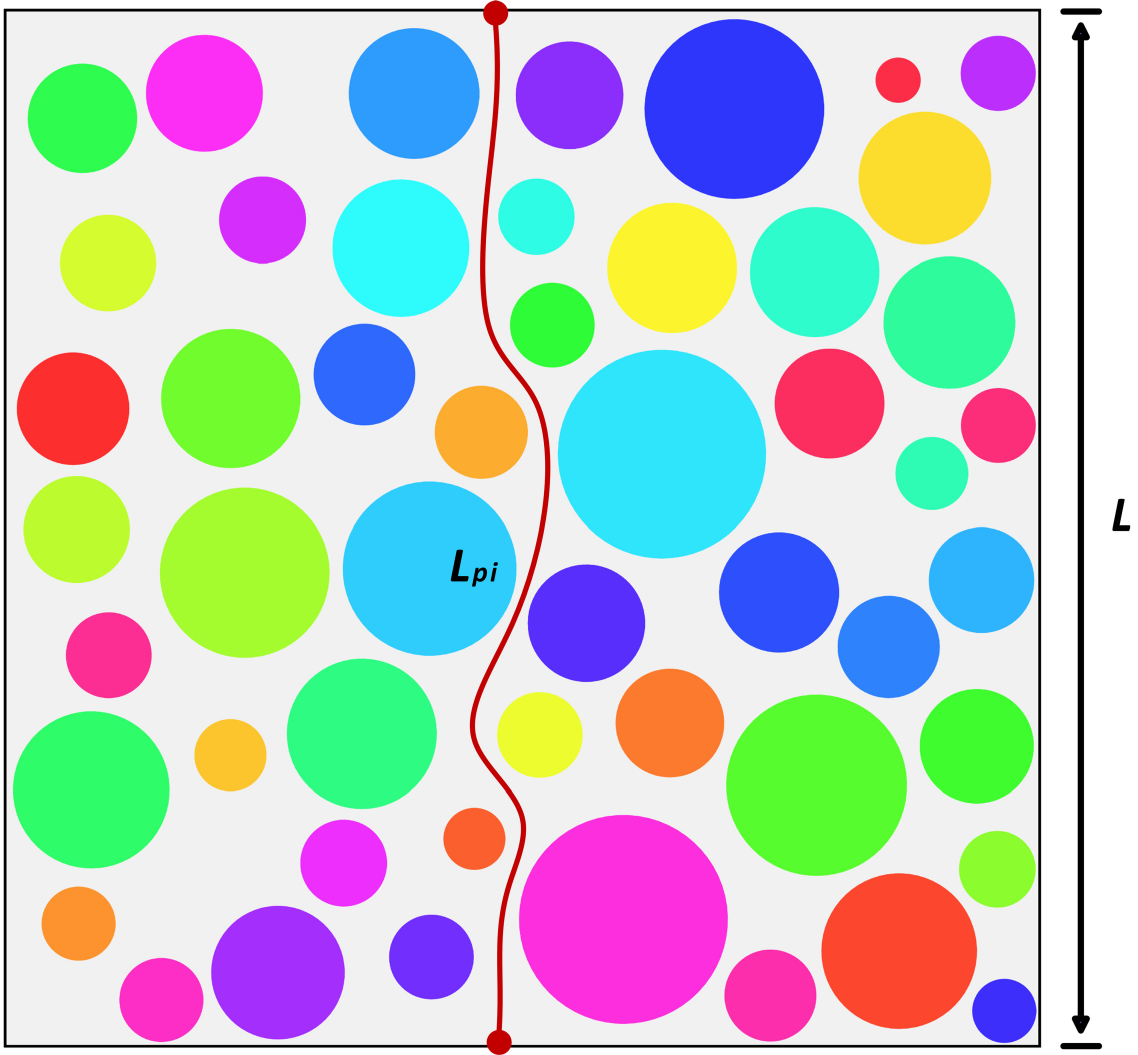}
   \caption{Schematic of a tortuous path of length $L_{pi}$ through a porous microstructure of thickness $L$, where the shortest tortuous path is used to calculate the tortuosity of the sample.}
   \label{fig:Fig1_Schematic_of_a_tortuous}
\end{figure}

\begin{figure}[h]
   \centering
   \includegraphics[scale=0.55]{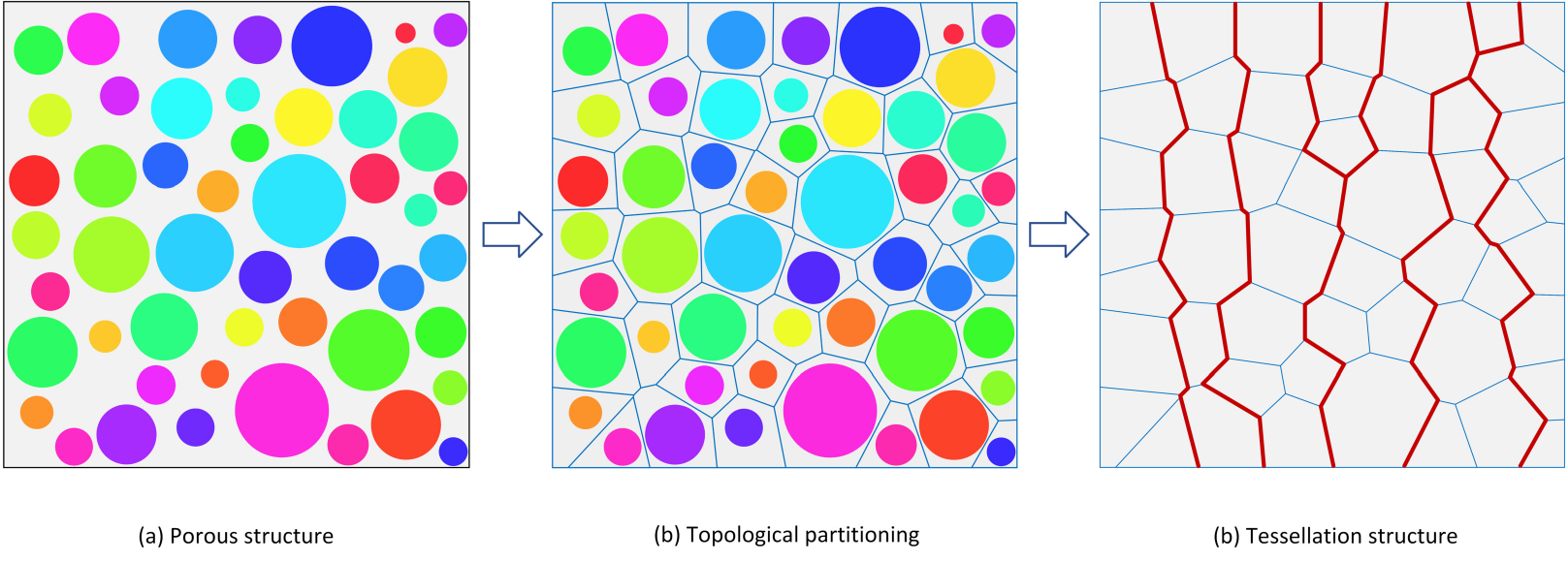}
   \caption{Schematic of the procedure for estimation of the diffusion path length in the porous structure, which includes mainly three steps: (a) generation of the porous structure with randomly packed particles; (b) spatial division of the porous packing structure; and (c) search for the shortest path. For clarity, only 2D case is illustrated, but it also applies to 3D.}
   \label{fig:Fig2_Schematic_of_a_procedure}
\end{figure}

As shown in \cref{fig:Fig1_Schematic_of_a_tortuous}, the tortuosity $\tau$ is defined as the ratio between the shortest Euclidean distance through a porous structure and the linear distance between the two edges of interest
\begin{align}
\tau  = \frac{\min \left\lbrace L_{pi}\right\rbrace}{L}
\label{eq:Eq1_equation_of_tortuosity}
\end{align}
which indicates that the value of $\tau$ is always equal to or greater than 1.
In order to find all possible paths through the pores, three procedures are performed, as shown in \cref{fig:Fig2_Schematic_of_a_procedure}: 
\begin{enumerate}[label=(\alph*)]
\item generating a porous structure model with randomly packed particles; 
\item topological division of porous packing structures based on radical tessellation; 
\item search for the shortest path along the edges of the polygons (2D) or polyhedra (3D).
\end{enumerate}
For procedure (a), we notice that the porous structure of battery electrodes are comprised of particle aggregates. Thus, the structure can be numerically constructed by particle packing method, which is based on our previous work~\citep{CHEN_2022_PT118002}. The details of the remaining two procedures are explained below.

\subsection{Topologically partition of porous packing structure}\label{Topologically partition}

\begin{figure}[h]
   \centering
   \includegraphics[scale=0.55]{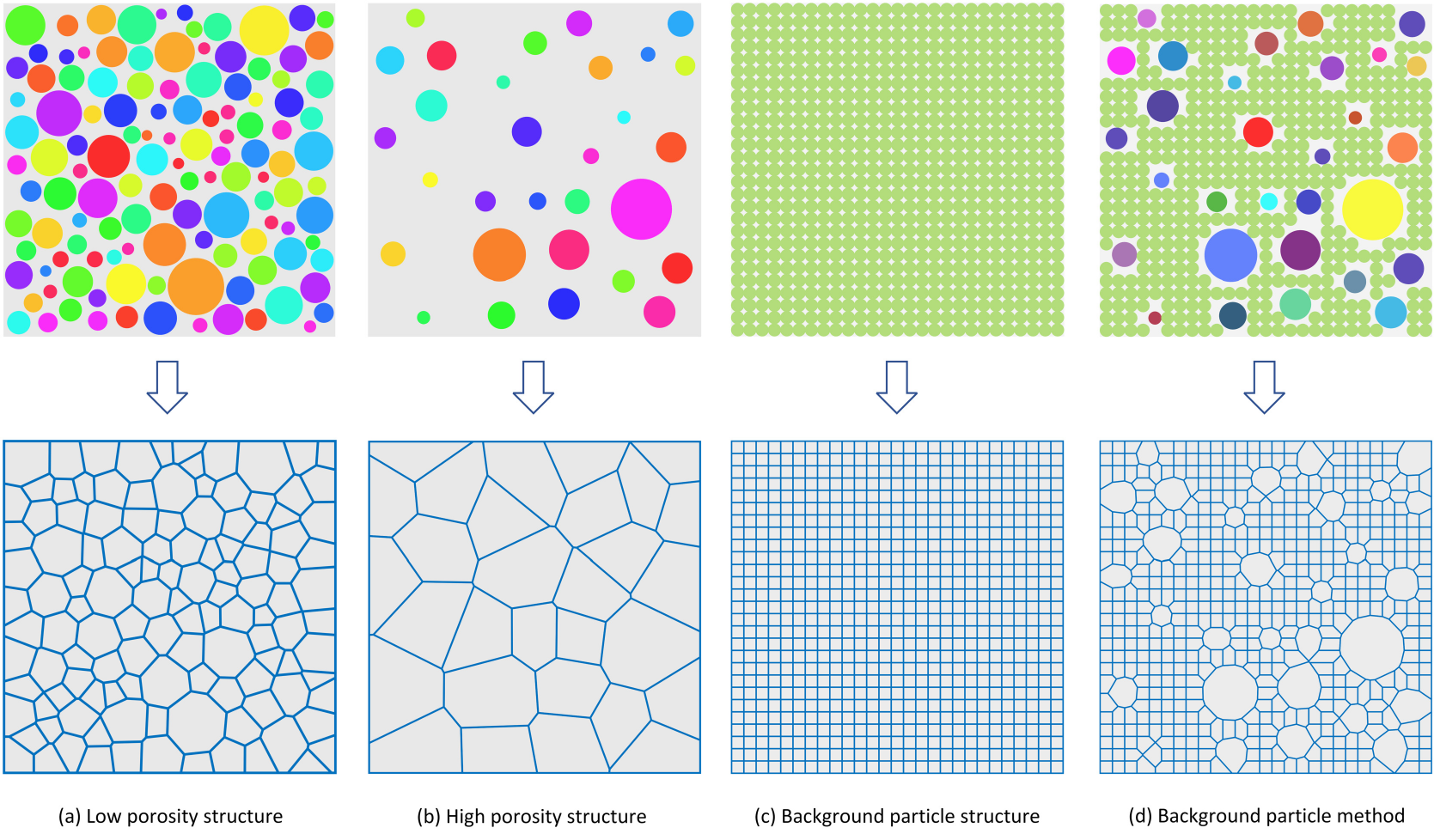}
   \caption{Schematic of the porous structures and their Radical tessellation for cases with (a) low porosity, (b) high porosity, (c) background particles structure, and (d) background particle method, which considers both the actual particles and the background particles.}
   \label{fig:Fig3_Schematic_of_RTP}
\end{figure}

For topological division of porous packing structures, the open-source program Voro++ developed by Rycroft is employed~\citep{Rycroft_2009_Chaos3215722}. For packing systems with low porosity (\cref{fig:Fig3_Schematic_of_RTP}a), in which case particles are densely packed and sufficiently enough cut lines are obtained after performing the topological partition, the estimation of tortuosity based on the original structure is satisfactory, and can agree well with values by other models. However, for loose packing systems with high porosity (\cref{fig:Fig3_Schematic_of_RTP}b)---in which case the pathways are mainly vertical and the tortuosity is close to 1---particles are located far away from each other and the cut lines are too slant to represent the real pathways. Therefore, the tortuosity estimated from this original structure is much higher than the real tortuosity. In order to remedy this situation, we introduce a group of fictitious background particles, assuming that the background particles are of equal size and are uniformly distributed at the vertices of a square grid in the computational domain, with the side length of the square equal to the diameter of the particles, as shown in \cref{fig:Fig3_Schematic_of_RTP}c. Since the background particles are of equal size and located at the corners of the squares (or cubes in 3D), the tortuosity of this structure is 1, ensuring the shortest pathway through the pore area with no obstacles in the way~\citep{Shi_2021_AEM03663}. For convenience, those particles that actually make up the porous structure, as opposed to fictitious background particles, are referred to as real particles. By overlaying the real particles domain on top of the background particle domain and removing the background particles that overlap with the real particles, we can effectively reproduce the diffusion path in the pore area based on topological partitioning of this structure and correctly estimate the tortuosity, as shown in \cref{fig:Fig3_Schematic_of_RTP}d. It should also be noted that, the size of the background particles is closely related to the computational accuracy and thus a careful determination of background particle size based on the real particle sizes should be conducted.

After obtaining all possible diffusion paths (the edges) in the porous packing structure, Dijkstra's path search algorithm is employed to find all possible paths from the starting point to the end point~\citep{Dijkstra_1959_NM86390,XU_2007_AMC094}.

\begin{figure}[h]
   \centering
   \includegraphics[scale=0.55]{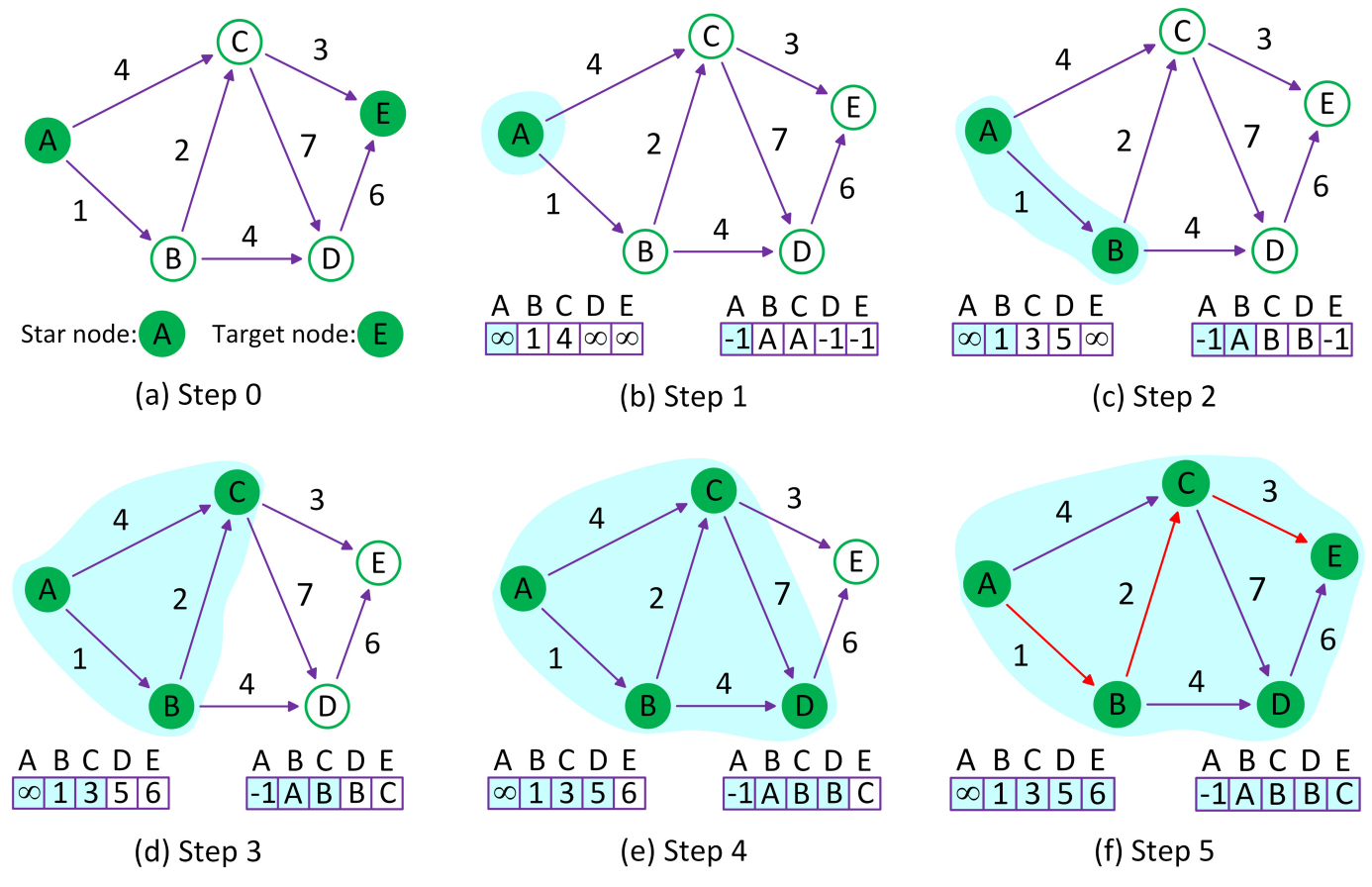}
   \caption{Schematic of Dijkstra's path search algorithm.}
   \label{fig:Fig4_Schematic_of_Dijkstra}
\end{figure}

Dijkstra’s path search algorithm is employed to find all possible paths from the starting point to the end point. Taking node $\text{A}$ as the start node and node $\text{E}$ as the target node, the aim is to find the shortest path from node $A$ through the rest of the nodes to reach node $\text{E}$, as shown in \cref{fig:Fig4_Schematic_of_Dijkstra}(a). Firstly, set up two arrays: the shortest distance array and the precursor array. If node $\text{A}$ is connected to node $\text{i}$, set the predecessor array $\text{P}(\text{A}, \text{i}) = 1$; otherwise, set $\text{P}(\text{A}, \text{i}) = -1$. Similarly, initialize the shortest distance array $\text{dist(A, i)}$ to the actual distance; otherwise, set $\text{dist(A, i)}$ to infinity. We start from the starting node $\text{A}$. It is known that node $\text{A}$ is connected to node $\text{B}$ and node $\text{C}$ respectively, so we can get the initial shortest distance array and precursor array, as shown in \cref{fig:Fig4_Schematic_of_Dijkstra}(b). In the shortest distance array, it can be observed that the shortest distance to reach node $\text{A}$ is 1, and its corresponding node is $\text{B}$. Therefore, the first step is to search node $\text{B}$ and its connected nodes. It can be observed that the next reachable nodes from node $\text{B}$ are node $\text{C}$ and node $\text{D}$. The total distance from node $\text{A}$ to node $\text{C}$ via node $\text{B}$ is 3, which is shorter than the direct distance from node $\text{A}$ to node $\text{C}$. At this point, the shortest distance array is updated with the value 3 for the distance from node $\text{A}$ to node $\text{C}$. The total distance from node $\text{A}$ to node $\text{D}$ via node $\text{B}$ is 5, so the shortest distance array is updated with the value 5 for the distance from node $\text{A}$ to node $\text{D}$. The precursor nodes of node $\text{C}$ and node $\text{D}$ are then updated to node $\text{B}$, as shown in \cref{fig:Fig4_Schematic_of_Dijkstra}(c). Next, in the shortest distance array, it can be observed that among the unsearched nodes, the shortest distance from node $\text{C}$ to node $\text{A}$ is 3, and its precursor node is $\text{B}$. Therefore, in this step, we search node $\text{C}$ and its adjacent nodes. It can be observed that the next reachable nodes from node $\text{C}$ are node $\text{D}$ and node $\text{E}$. Here, the total distance from node $\text{A}$ to node $\text{D}$ through nodes $\text{B}$ and $\text{C}$ is 10, which is greater than the distance from node $\text{A}$ to node $\text{D}$ through node $\text{B}$. Therefore, the shortest distance array value for the distance from node $\text{A}$ to node $\text{D}$ and its precursor node are not updated. However, the total distance from node $\text{A}$ to node $\text{E}$ through nodes $\text{B}$ and $\text{C}$ is 6. The shortest distance array value for the distance from node $\text{A}$ to node $\text{EC}$ is updated to 6, and the precursor of node $\text{E}$ is updated to node $\text{C}$, as shown in \cref{fig:Fig4_Schematic_of_Dijkstra}(d). The same process is repeated until the target node $\text{E}$ is reached. Finally, the shortest path from node $\text{A}$ to node $\text{E}$ is found to be $\text{A-B-C-E}$, with a path length of 6, as shown in \cref{fig:Fig4_Schematic_of_Dijkstra}(f).

In practice, the start node at one side of the porous packing structure is denoted by $s_i$, and the corresponding target node at the other side is denoted by $t_i$. The shortest path $\mathrm{dist}(s_i,t_i)$ obtained by the Dijkstra path search algorithm, the tortuosity $\tau_{s_i,t_i}$ corresponding to each vertex, and the mean tortuosity $\bar\tau$ can be calculated using the following equations:
\begin{align}
&{\tau _{{s_i},{t_i}}} = \frac{\mathrm{dist}({s_i},{t_i})}{L}\\
&\bar{\tau} = \frac{1}{{N}} \sum\limits_{i = 1}^N \tau_{{s_i,t_i}}
\end{align}
where $N$ is the total number of vertex pairs.

\begin{figure}[H]
  \centering
  \subfigure[Selected porous media domains.]{
    \centering
    \includegraphics[scale=0.36]{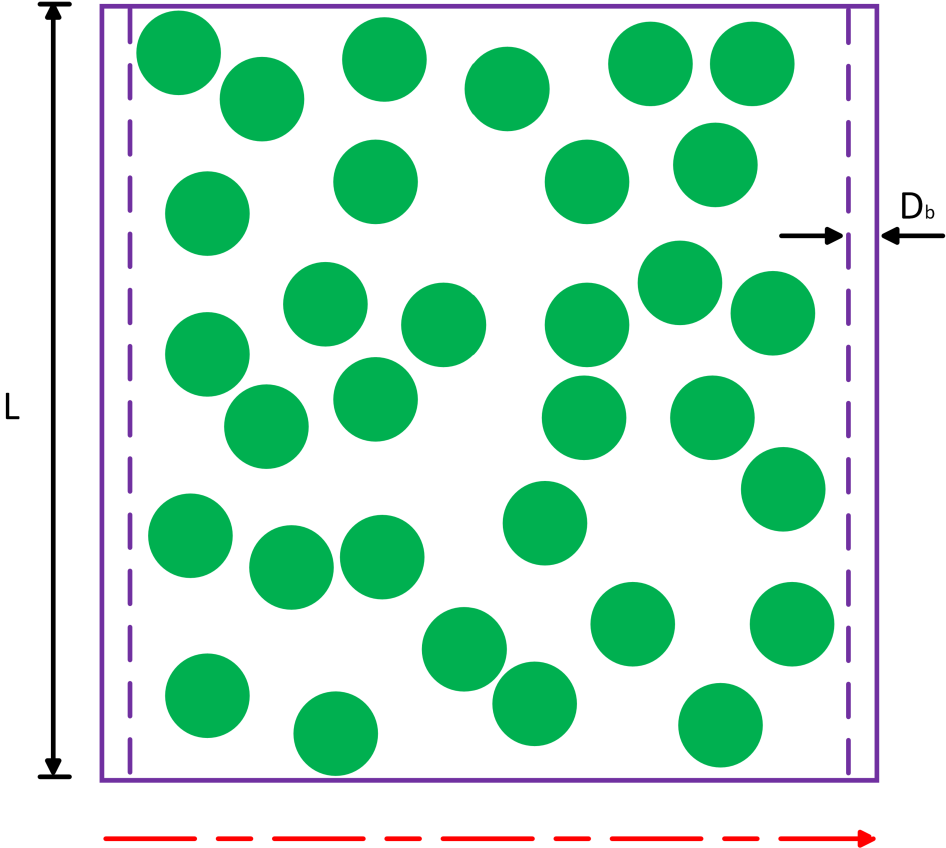}\label{fig:Fig5a_Selected_porous_media_domains}
  }
\hspace{1.0cm}
    \subfigure[Porous structure with 2 lines of background particles.]{
    \centering
    \includegraphics[scale=0.36]{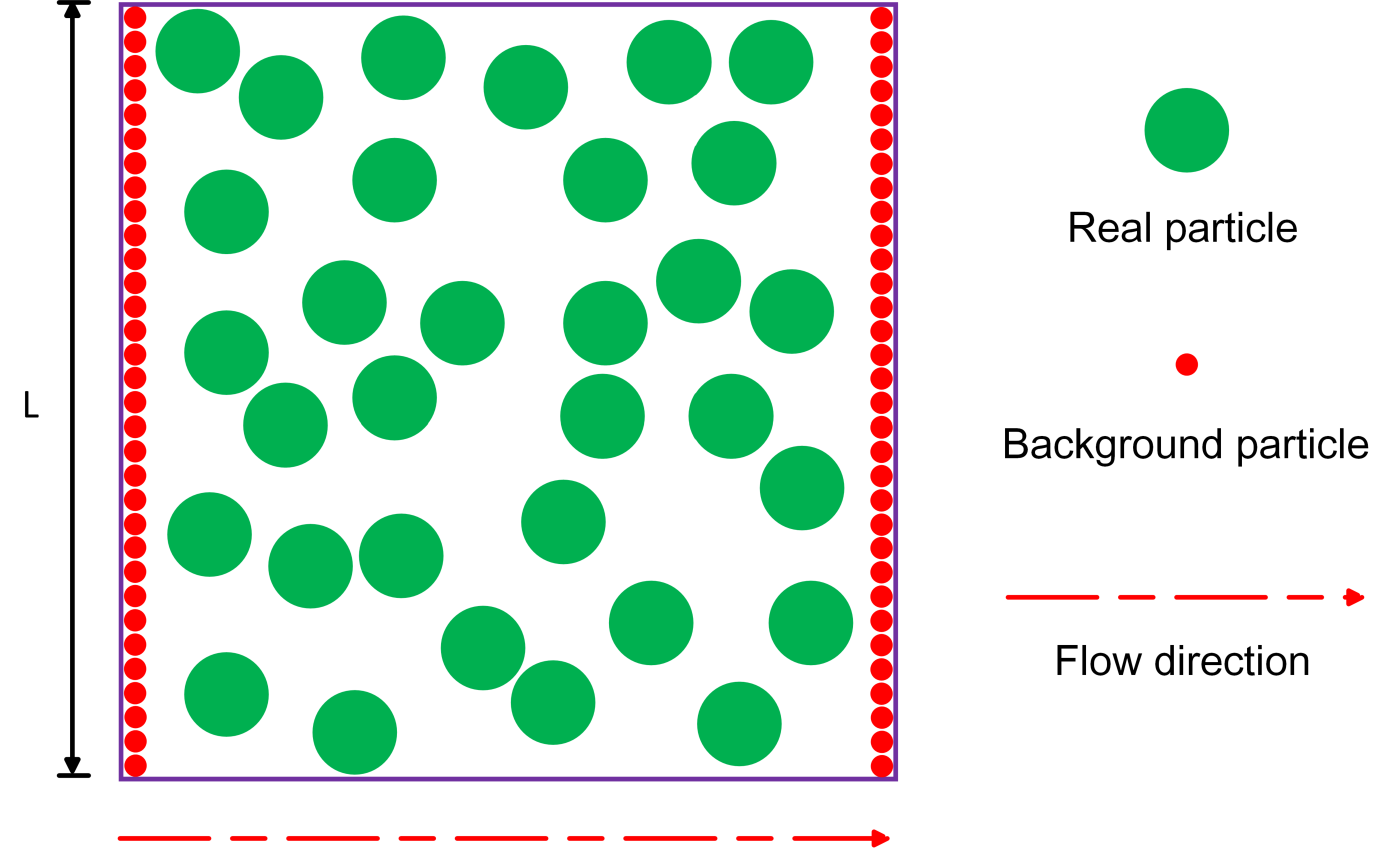}\label{fig:Fig5b_Structure_of_DM}
  }
 \caption{Schematic of the (a) porous structure and (b) porous structure with 2 lines of background particles as in- and out-let of the pathways.}\label{fig:Fig5_Model_Structure}
\end{figure}

Furthermore, in order to ensure that the starting node can be exactly paired with a target node in the route search algorithm, a column of background particles in each inlet and outlet boundary for all cases is placed, as shown in \cref{fig:Fig5_Model_Structure}. Therefore, the distance of each particle of interest from the boundaries $D_b$ must not be smaller than the size of the background particles. In this way we can make sure that, the pathways from topological division based on the original configuration and that based on the structure with background particles, have consistent starting and ending points for the search. Such consistency helps to eliminate differences in the initial conditions and provides a basis for subsequent quantitative comparisons of the computational accuracies of the two methods, which is essential for evaluating the advantages and disadvantages of the two methods. 

\subsection{The method of particle aggregation morphology recognition}\label{sec:Aggregation morphology recognition}
\begin{figure}[H]
  \centering
  \subfigure[Packing structure.]{
    \centering
    \includegraphics[scale=0.33]{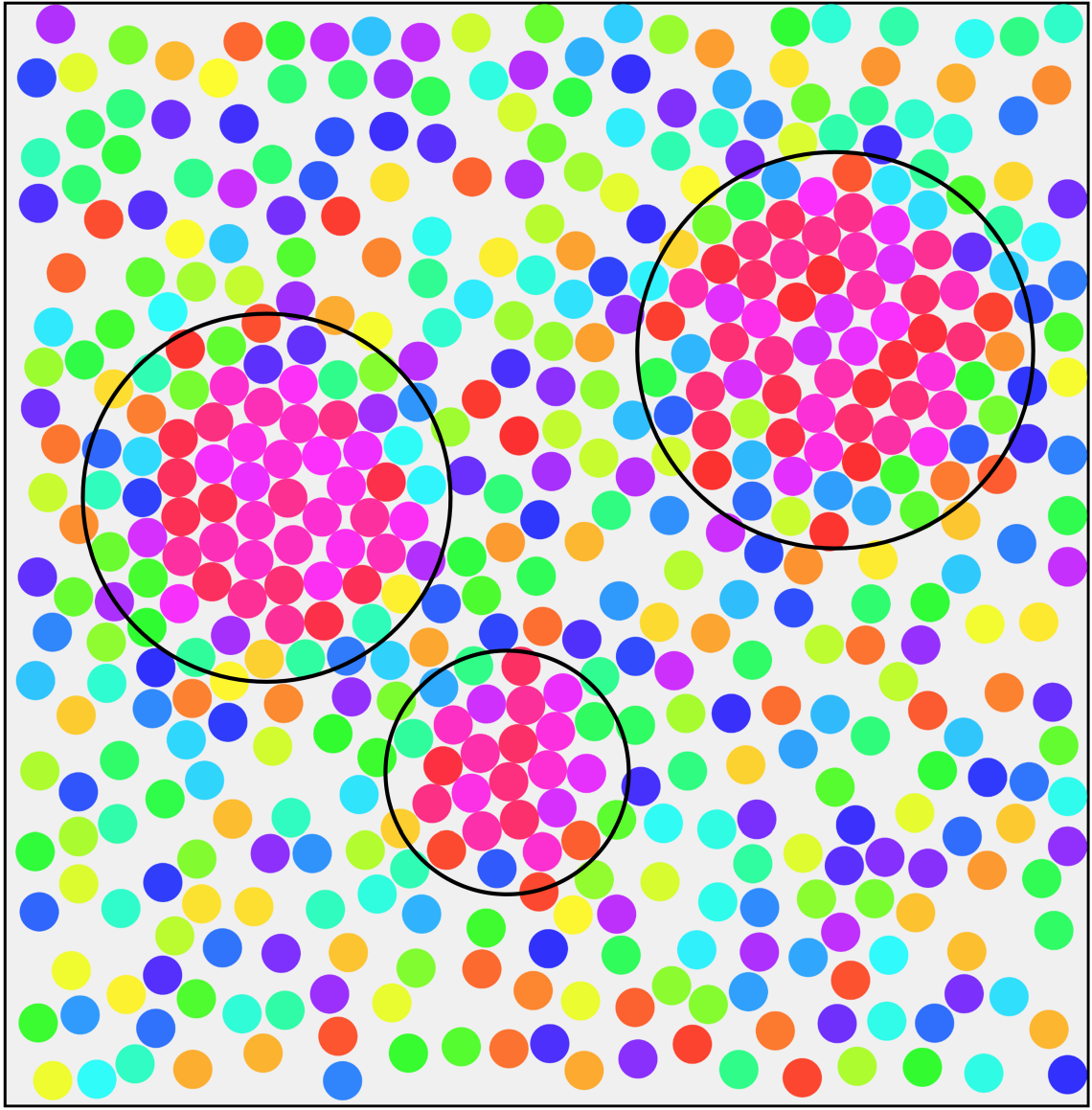}\label{fig:Fig4a_packing_structure}
  }
\hspace{1.0cm}
    \subfigure[Aggregation morphology recognition.]{
    \centering
    \includegraphics[scale=0.33]{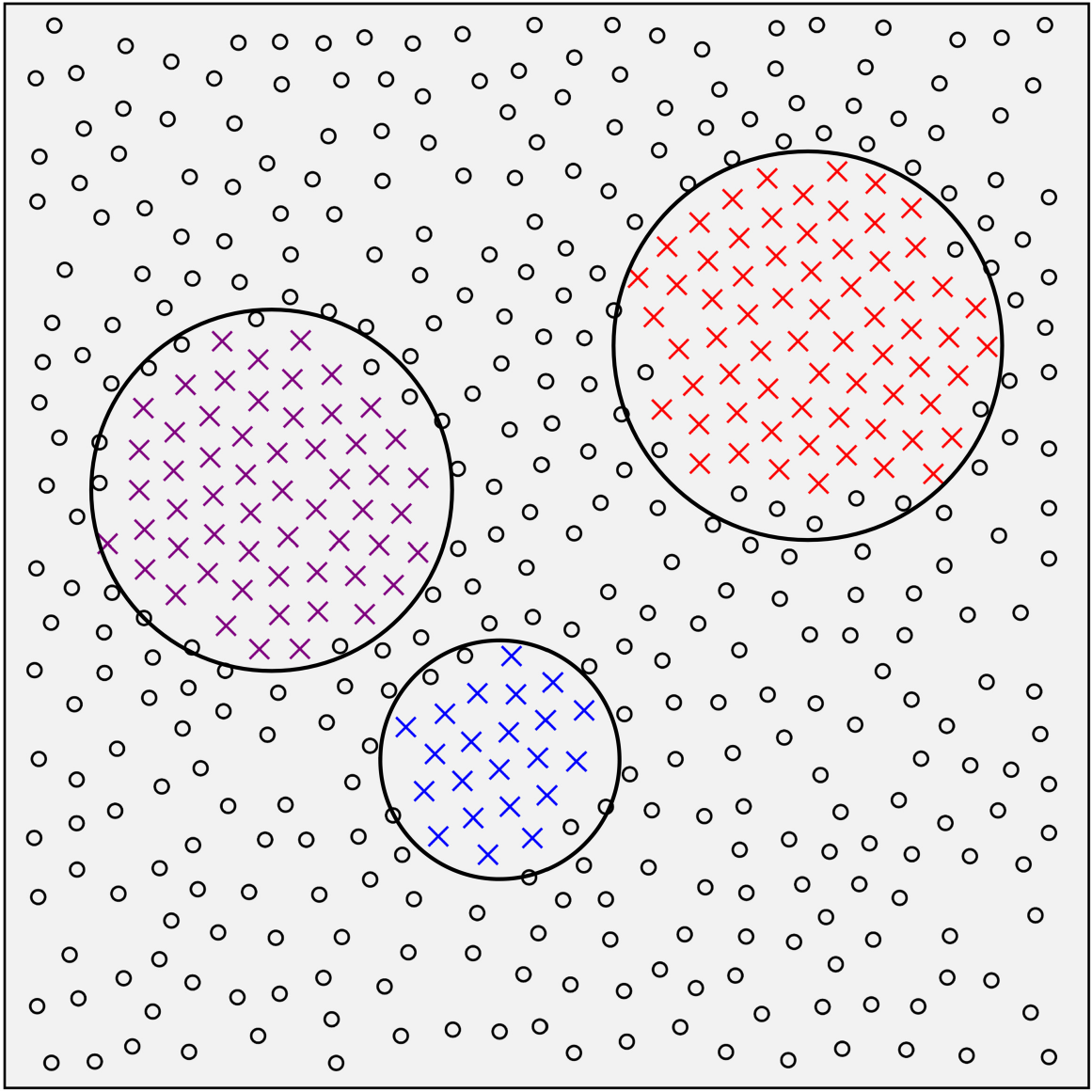}\label{fig:Fig4b_morphology_recognition}
  }
 \caption{Example of particle aggregation morphology recognition.}\label{fig:Fig4_Example_aggregation_morphology}
\end{figure}

The aggregation of particles has a significant impact on the morphology of porous structures, thus affecting their tortuosity. Unlike porous particle packing structures without aggregation, we intentionally place particles at specific locations during the initial particle generation process, which leads to significant overlap between particles. By employing the method developed in our previous work to move these particles, we achieve the objective of creating porous particle packing structures characterized by particle aggregation~\citep{CHEN_2022_PT118002}. The degree of particle aggregation is generally characterized by the the size and quantity of the aggregates. In this work, Density-Based Spatial Clustering of Applications with Noise (DBSCAN) method is employed for for the evaluation of particle aggregation, which is a density-based clustering algorithm~\citep{Ester_1996_AAA,HANAFI_2022_ESA117501}. In particle aggregation assessment, DBSCAN can be used to identify areas of dense particles and to help determine the size and number of aggregates, with excellent recognition of aggregates of any shape and size.   \Cref{fig:Fig4_Example_aggregation_morphology} shows an example of aggregation morphology recognized by DBSCAN.

The aggregation index of particles in a porous structure can be calculated using the following equation:
\begin{align}
\text{Aggregation Index} = \frac{{V_{\mathrm{aggregation}}}}{{V_{\mathrm{total}}}}
\end{align}
where, $V_{\mathrm{aggregation}}$ denotes the total volume of particles involved in aggregation, and $V_{\mathrm{total}}$ denotes the total volume of the particles.

\section{Results and discussions}

In this section, the variation of tortuosity of porous structures with different morphologies is discussed. Specifically, porous media with different particle sizes and volume fractions are constructed based on the particle packing algorithm which proposed in our previous work~\citep{CHEN_2022_PT118002}, and the tortuosity of pore structures with different morphologies is estimated using the method described in Section 2. 

%

\subsection{Model validation}\label{sec:model_validation}
\begin{figure}[H]
  \centering
  \subfigure[FEM simulation.]{
    \centering
    \includegraphics[scale=0.35]{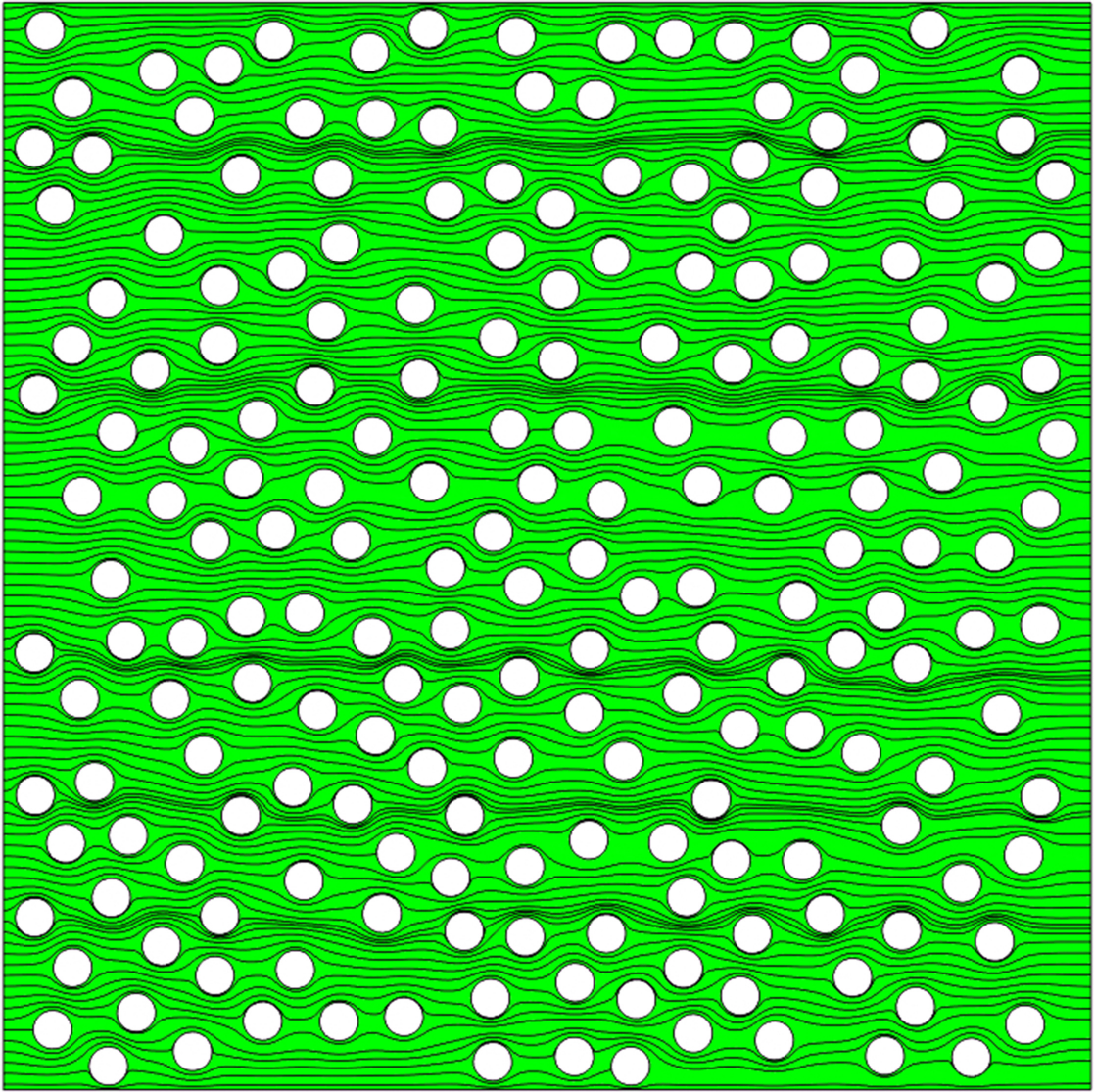}\label{fig:Fig6a_FEM_electric_flow}
  }
\hspace{1.0cm}
    \subfigure[Radical tessellation based method.]{
    \centering
    \includegraphics[scale=0.5901]{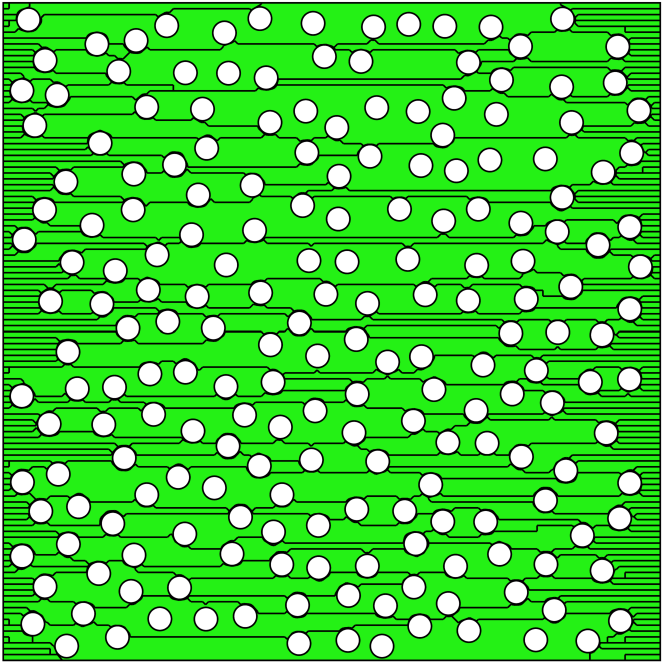}\label{fig:Fig6b_electric_flow}
  }
 \caption{Streamlines of diffusional paths.}\label{fig:Fig6_electric_flow}
\end{figure}

\begin{figure}[h]
   \centering
   \includegraphics[scale=0.45]{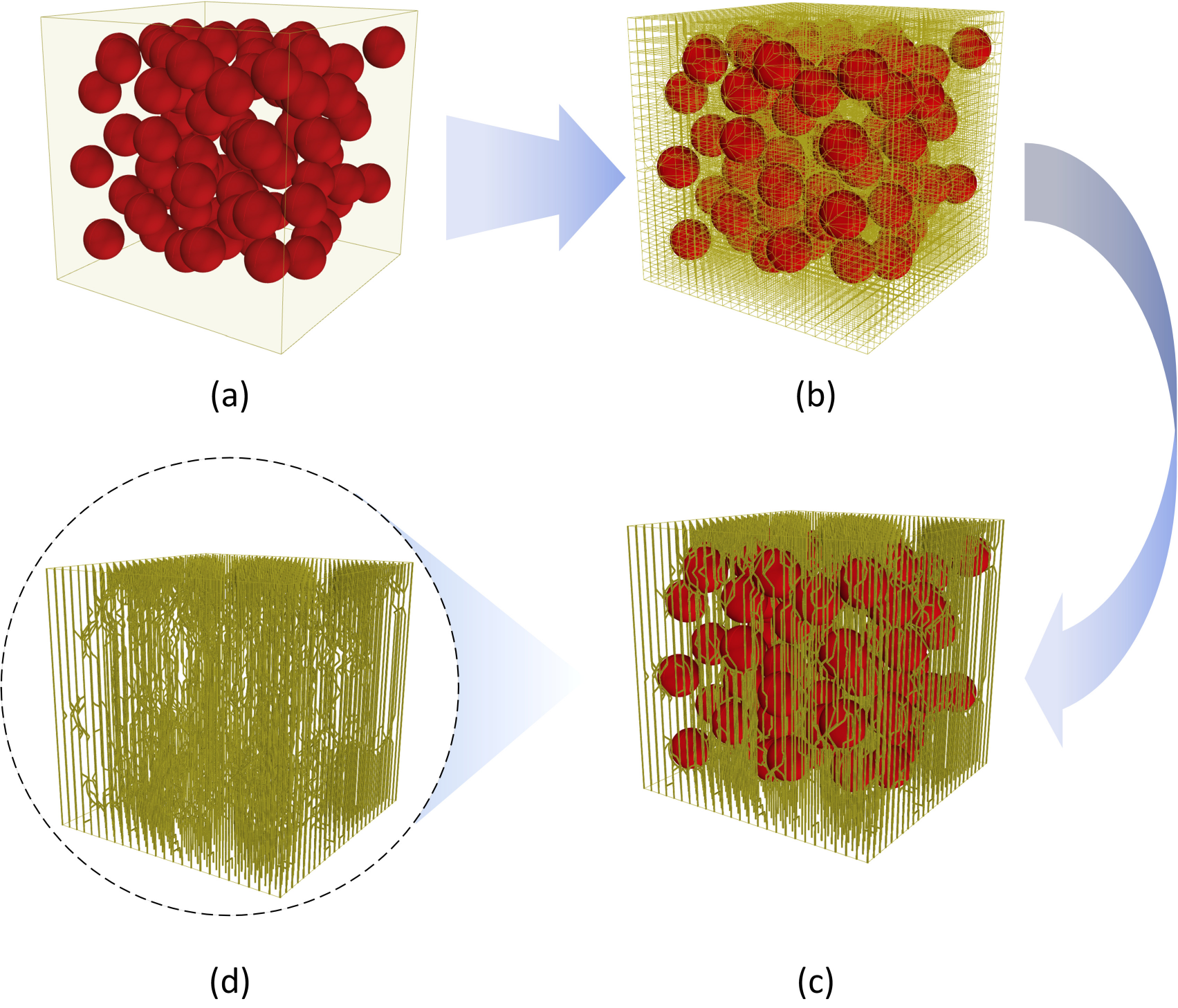}
   \caption{Schematic of identification of diffusion pathways through a 3D porous structure by tessellation-based method: (a) The packing structure; (b) The skeleton extracted from the packing structure for tessellation; (c) The shortest pore paths identified between inlet and outlet.(d) The view of paths without the particles.}
   \label{fig:Fig7_Schematic_of_3D_simulations}
\end{figure}

\begin{figure}[H]
  \centering
  \subfigure[Model validation, 2D.]{
    \centering
    \includegraphics[scale=0.33]{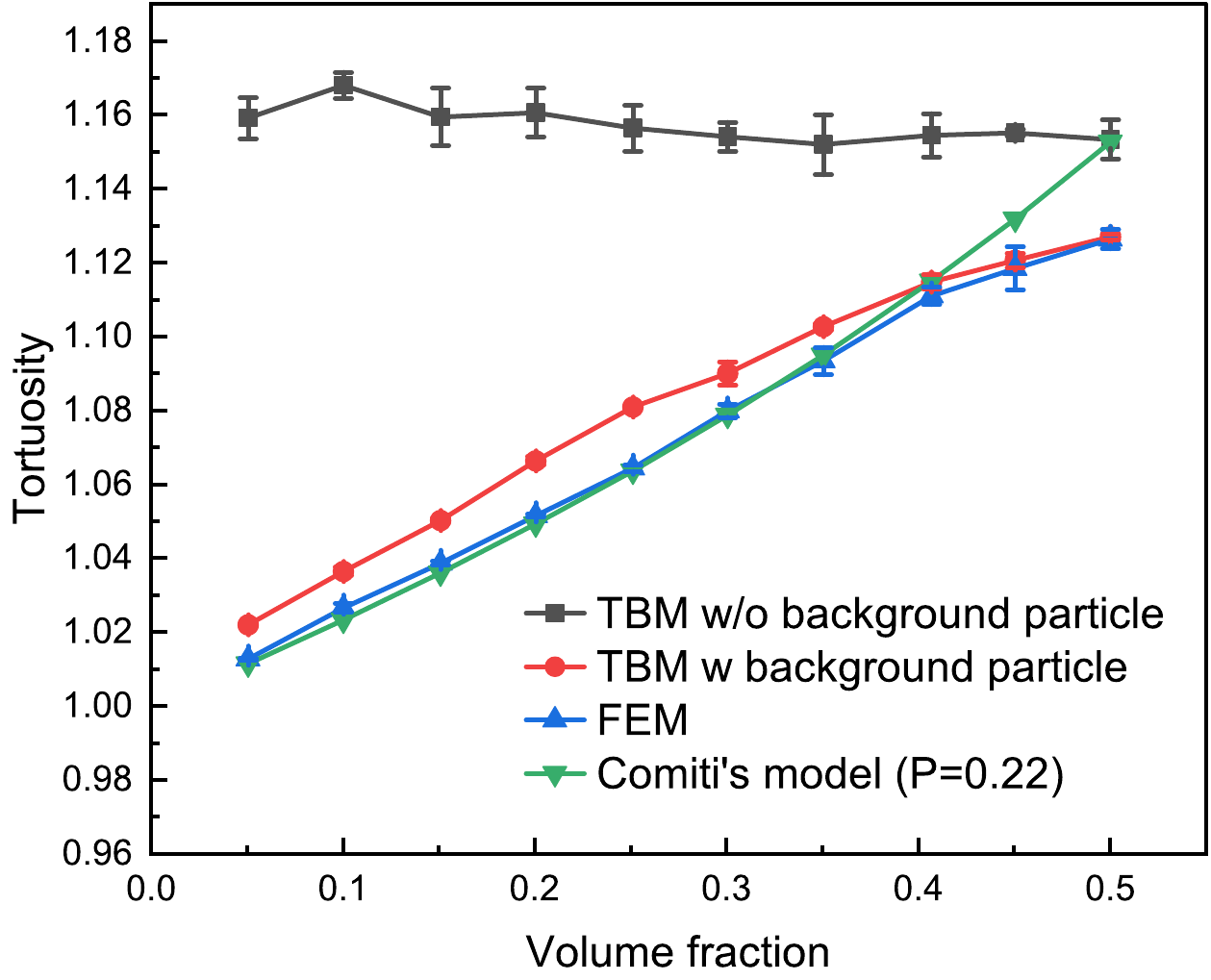}\label{fig:Fig8a_Compare_FEM}
  }
\hspace{1.0cm}
    \subfigure[Model validation, 3D.]{
    \centering
    \includegraphics[scale=0.33]{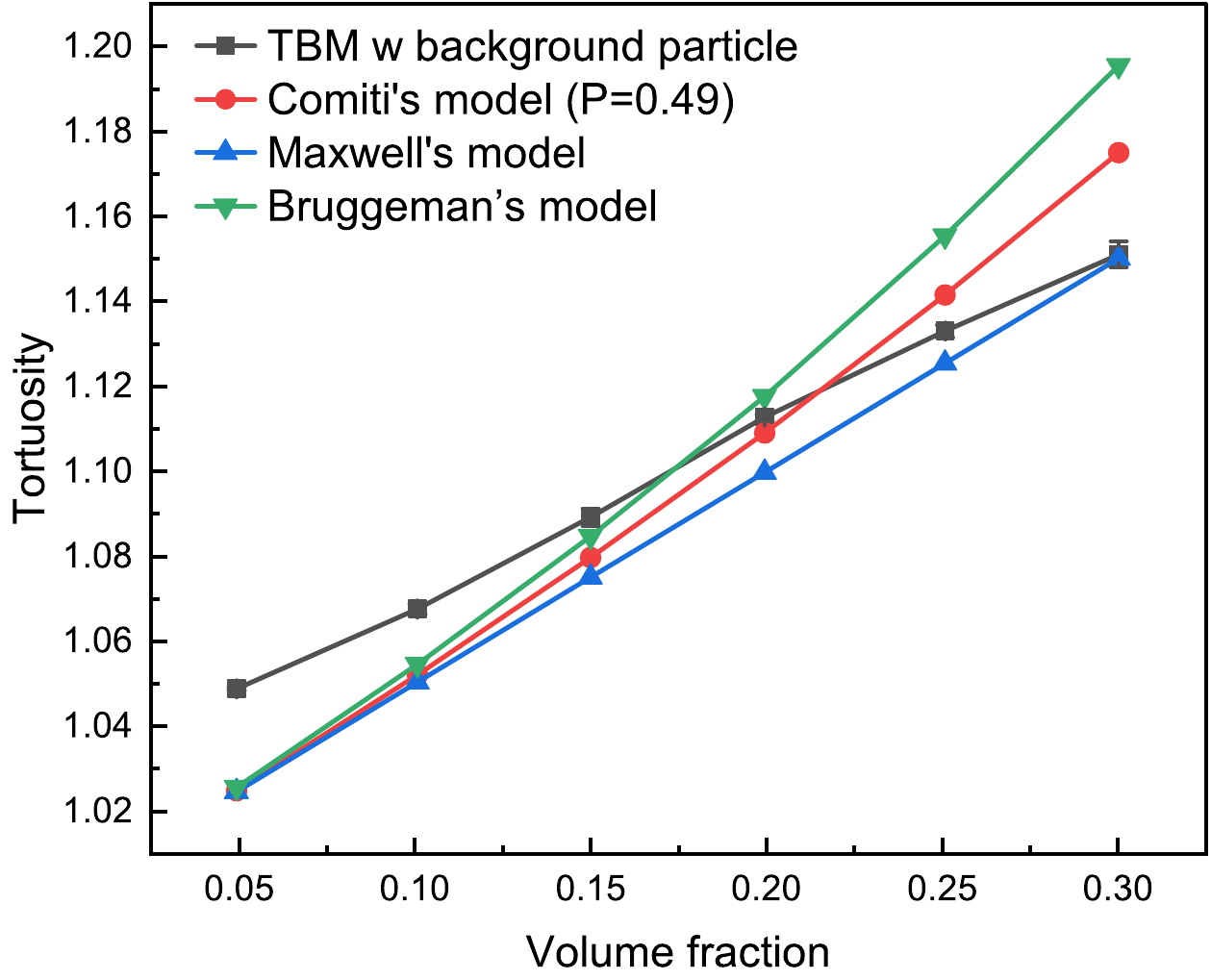}\label{fig:Fig8b_Compare_EMP}
  }
 \caption{Comparison of the tortuosity estimation of different methods.}
 \label{fig:Fig8_Compare_FEM_EMP}
\end{figure}

First, we compare our tessellation-based method with the finite element method (FEM). Since we do not need to account for time variations in estimating the diffusion tortuosity, a steady-state diffusion process is simulated, whose governing equation is the Laplace equation
\begin{align}
&\bm\nabla^2 V = 0
\end{align}
in the computational domain. As for the boundary conditions, we set $V=0$ at the inlet boundary, $V=1$ at the outlet boundary, and flux-free condition $\bm\nabla V\cdot \bm n=0$ for the other boundaries. The simulation domain is a square with $L\times L= 225\times 225$,  the particles are equal sized with a radius of $R=4$. Also, in order to avoid dense FE mesh due to particles contacting each other, the distances between two particles are set to be greater than a certain value $R_\varepsilon$ (in this work $R_\varepsilon=0.1R$ with $R$ being the real particle radius). We also employed the tessellation-based method to calculate the diffusion paths for the same structure, using fictitious background particles with a radius of $r=1$ as comparison. The streamlines of the diffusion paths obtained by FE simulations and the tessellation-based method in a square of  are shown in \cref{fig:Fig6a_FEM_electric_flow} and \cref{fig:Fig6b_electric_flow}, respectively. It is shown that for a given structure, despite differences in path smoothness and quantity, the results of both methods are consistent in terms of overall trends and directions. Therefore, in this case, the differences in paths do not significantly affect the tortuosity estimation results. Since the paths from tessellation-based method is constructed by edges of polygons, they naturally have more corners than those from FE simulations. As for the difference of number of paths, it can be explained by the fact that the algorithm of tessellation-base method only allows the search for optimal path, i.e. shortest path, to be chosen, omitting all the other possible paths. FE simulations, on the other hand, calculate diffusion paths based on global optimization of the energy, taking interaction of different paths into account, thus offering multiple diffusion paths alone the same path.
Although it is probably more physically sound to use FEM for the calculation of tortuosity, the computational consumption is way much greater than the tessellation-based method. Moreover, it is even more difficult for FEM to obtain pathways in 3D structure with a group of several dozen particles, but can be easily done by the tessellation-based method, as shown in \cref{fig:Fig7_Schematic_of_3D_simulations}. The same as in the 2D case, the skeleton structure of the pores in the packing structure is obtained based on topological segmentation, and then Dijkstra's algorithm is applied on the skeleton to identify the shortest paths, which can clearly capture the diffusion paths inside the pores.
In terms of quantitative comparison, the tessellation-based method yields results nearly identical to those of the FEM (about 1.4$\%$ higher in 2D) and the empirical models (about 1.6$\%$ higher in 2D) presented in \cref{tb:empirical_models}, both in 2D and 3D, as shown in \cref{fig:Fig8_Compare_FEM_EMP}. \Cref{fig:Fig8a_Compare_FEM} shows the comparison of tessellation-based method with FEM and Comiti's model (Eq. (\ref{model:1})). We can observe that the overall trend of tortuosity is nearly identical across the three methods, with the tessellation-based method yielding slightly higher values than the other two at lower volume fractions of real particles (about 1.4$\%$ higher than FEM and about 1.6$\%$ higher than Comiti's model in 2D). However, when the volume fraction of real particles increases, the tessellation-based method aligns more closely with FE simulations., outperforming the empirical method. It is also important to note that this improved accuracy is achieved through the introduction of fictitious background particles. If there is no background particles, we can not correctly estimate tortuosity, especially with loosely packed structure (low volume fraction). We can also compare the tortuosity estimated by different models in 3D, as shown in \cref{fig:Fig8b_Compare_EMP}. Since the FE simulations in 3D require too much computational resources than we can afford, we did not compare our results with those from FEM. Instead, comparison is performed with three empirical models shown in \cref{tb:empirical_models}. The results are nearly identical again, with slightly higher tortuosity at low volume fractions and lower tortuosity at high volume fractions. Specifically, it is about 2.4$\%$ higher at low volume fractions and about 2$\%$ lower at high volume fractions compared to Comiti's model. It is worth mentioning that most of the current commonly used tortuosity estimation models perform tortuosity estimation in 2D or 3D, but rarely both~\citep{FU_2021_ESR103439,Ghanbarian_2013_SSSAJ0435}. However, from \cref{fig:Fig8_Compare_FEM_EMP}, it can be found that the method proposed here can offer satisfactory results with affordable computational costs in both 2D and 3D. In this regard, the method proposed in this article undoubtedly has better applicability.

\begin{figure}[H]
   \centering
   \includegraphics[scale=0.45]{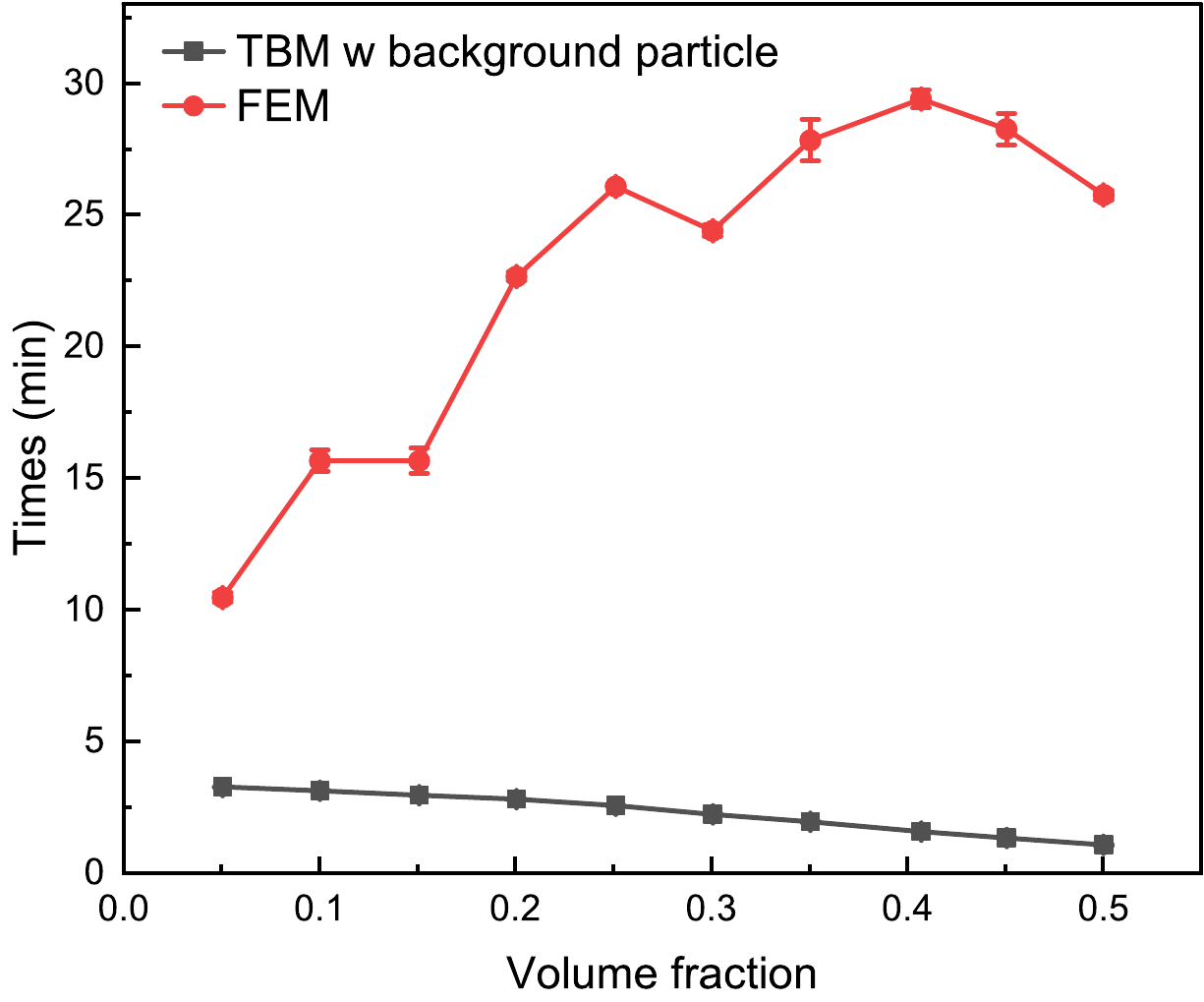}
   \caption{Comparison of the time consumption of different methods.}
   \label{fig:Fig9_time_coms}
\end{figure}

In order to have a more intuitive understanding of the computational efficiency of the algorithm proposed in this paper, the computational time consumed by the background particle method is compared with the time consumed by the finite element simulation in 2D. The time compared here sums up all that consumed in the procedures, including the topological division of porous media, geometric modeling, FE meshing and calculation. All computations are performed on a standard desktop computer equipped with an Intel Core i7-10700 processor, featuring eight CPU cores, a main CPU frequency of 2.9GHz, and 16GB of RAM. The comparative results are shown in \cref{fig:Fig9_time_coms}. It can be observed that the computational time required by the method in this paper is much smaller than that required for finite element simulation. As the volume fraction increases, the time consumption of finite element simulations shows a growing trend. In contrast, the time required for the method proposed in this article gradually decreases. The variation of computational efficiency is affected by multiple factors. On the one hand, in the TBM method, the number of background particles gradually decreases with the increase of the actual particle volume fraction, which significantly reduces the number of nodes after topological division. Specifically, the number of nodes to be searched in the path search process is reduced, which accelerates the search process and gradually reduces the required computation time. In contrast, in finite element simulation, as the particle volume fraction increases, the distance between particles gradually decreases, leading to the formation of numerous narrow gaps in the pore structure, which require a higher density mesh for processing. The higher mesh density not only increases the time required for mesh generation, but also significantly raises the complexity of the numerical solution process. In addition, it can also be observed that, compared to the finite element method, the tessellation-based method with background particles exhibits better computational time stability. The occurrence of this phenomenon can be explained by the background particles. In the TBM method, most of the topological division nodes come from background particles, which are less variable in the system with the same volume fraction, thus avoiding significant changes in computational time with changes in the actual particle locations. In contrast, the position changes of the particles can cause small regions in the system to shift, and the presence of these narrow faces has a significant impact on the mesh delineation, further leading to fluctuations in computational time. In addition, it can also be observed that, compared to the finite element method, the tessellation-based method with background particles exhibits better computational time stability. In summary, the TBM with pseudo background particles has a good performance in terms of computational accuracy, time consumption and stability of computational results.

\subsection{Effect of background particle size and thickness of the domain}
\begin{figure}[H]
   \centering
   \includegraphics[scale=0.45]{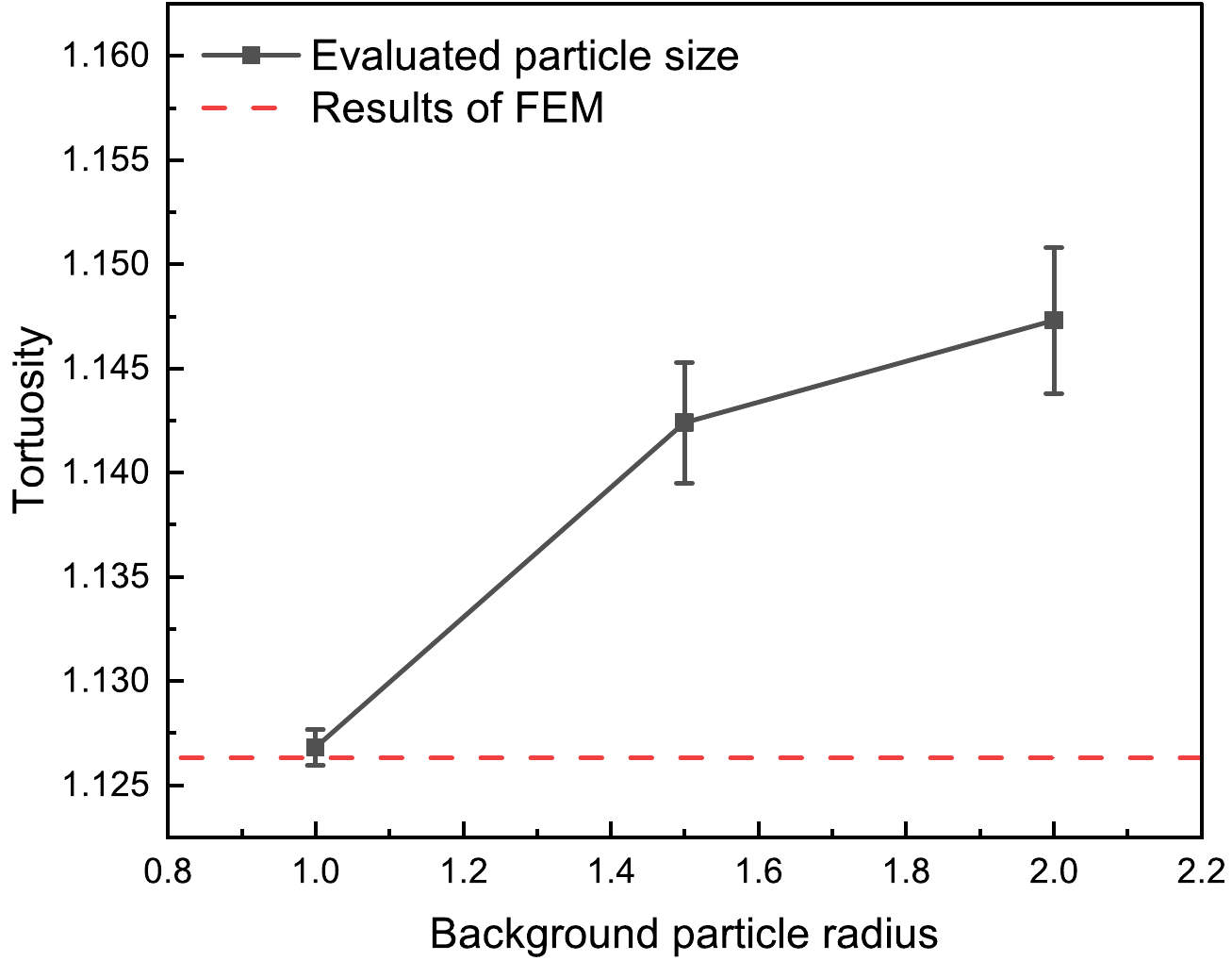}
   \caption{Effect of background particle size.}
   \label{fig:Fig10_particle_size}
\end{figure}

In \cref{sec:model_validation}, we demonstrated that TBM with pseudo background particles is effective in improving the reliability of the tortuosity estimation. However, it also raises the question: how susceptible is the estimation accuracy to the size of the background particles. In order to study the effect of particle size of  background particles on the accuracy of tortuosity estimation, three different ratios of radii between background and real particles are selected for comparison: $r/R=1/4$, $r/R=1.5/4$ and $r/R=2/4$. The real particles have a radii of $R=4$, and the volume fraction of the real particle is $50\%$. Referring to the sample size used by Espinoza-Andaluz et al.\citep{Mayken_2020_MCS017}, for each case, nine structures are randomly generated and analyzed statistically, as shown in \cref{fig:Fig10_particle_size}. It can be observed that as the background particle size increases, the estimated tortuosity becomes higher. From \cref{fig:Fig8_Compare_FEM_EMP} we know that for the case in this section, the tortuosity is approximately $\tau=1.1263$. Therefore, we can further conclude that the estimation accuracy of tortuosity decreases continuously with the increase of background particle size. Moreover, as the particle size decreases, the error bar becomes significantly smaller, indicating more reliable estimation of tortuosity. However, it is should also be noted that excessively small background particles, while offering higher accuracy in estimating tortuosity, will increase the computational cost. Therefore, selecting an appropriate background particle size is crucial for the computational efficiency of the tortuosity estimation results of our method. From our calculation, it is suggests that when the diameter of background particles is less than or equal to 1/4 of the particle size that makes up the porous structure, we can achieve an optimized balance between the accuracy of tortuosity estimation and computational efficiency.

\begin{figure}[H]
   \centering
   \includegraphics[scale=0.45]{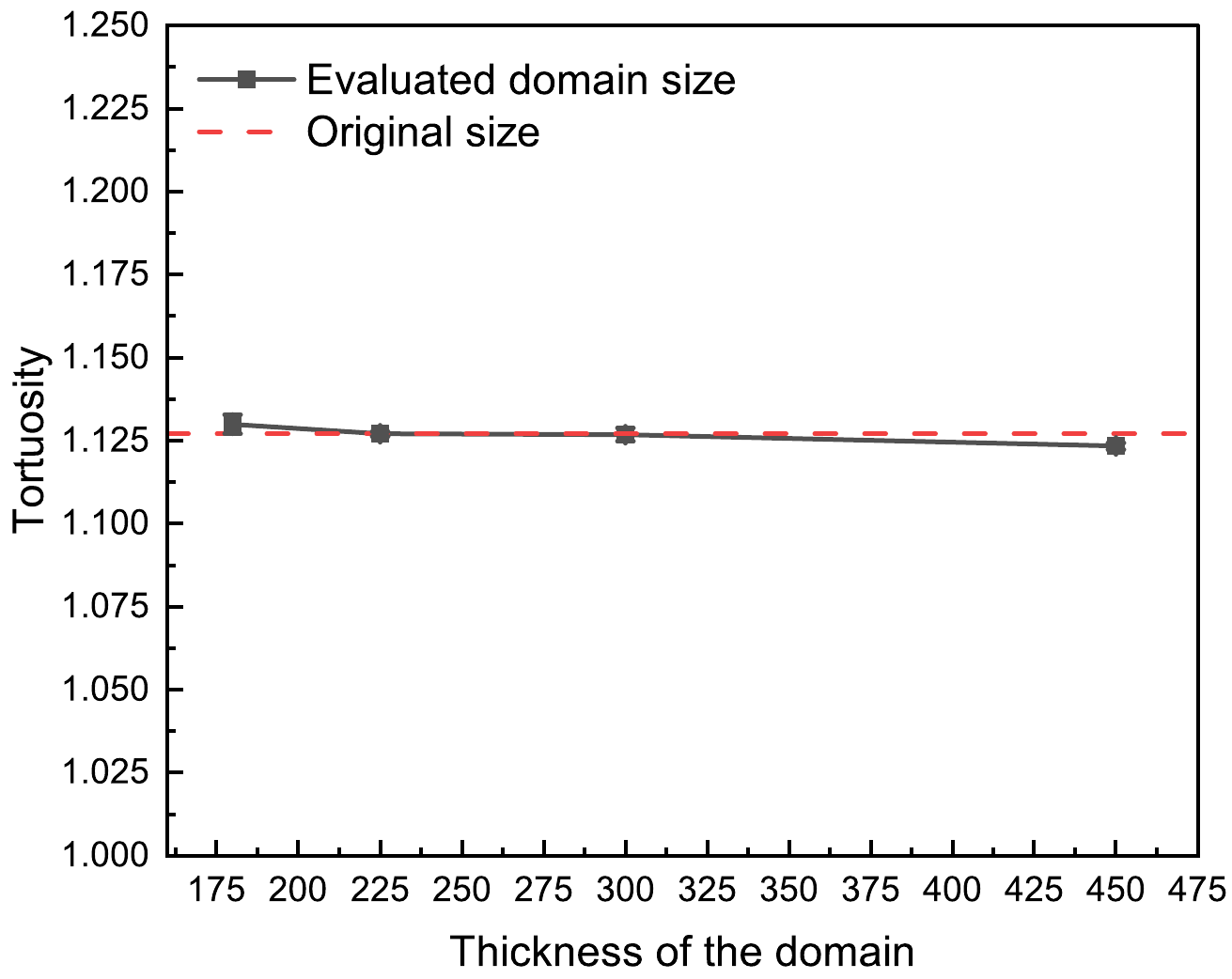}
   \caption{Effect of domain size.}
   \label{fig:Fig11_domain_size}
\end{figure}

Now we discuss the influence of simulation domain size on the tortuosity estimation. The domain sizes investigated range from 225 to 450 and all other parameters, including particle size and volume fraction, remain the same as in \cref{fig:Fig10_particle_size}. The result is plotted in \cref{fig:Fig11_domain_size}, which shows that, as the computational domain expands, the value of tortuosity may undergo slight variations, but the difference in its values is not significant (within $0.1\%$). Therefore, it can be concluded that the size of the computational domain has little effect on the value of tortuosity when the particle size is consistent. This conclusion is consistent with the research findings of Espinoza Andaluz et al., whose research also proves that the difference in computational domain has no significant impact on the value of tortuosity~\citep{Mayken_2020_MCS017}. 

\subsection{Effect of porosity and particle size on tortuosity}
\begin{figure}[H]
   \centering
   \includegraphics[scale=0.45]{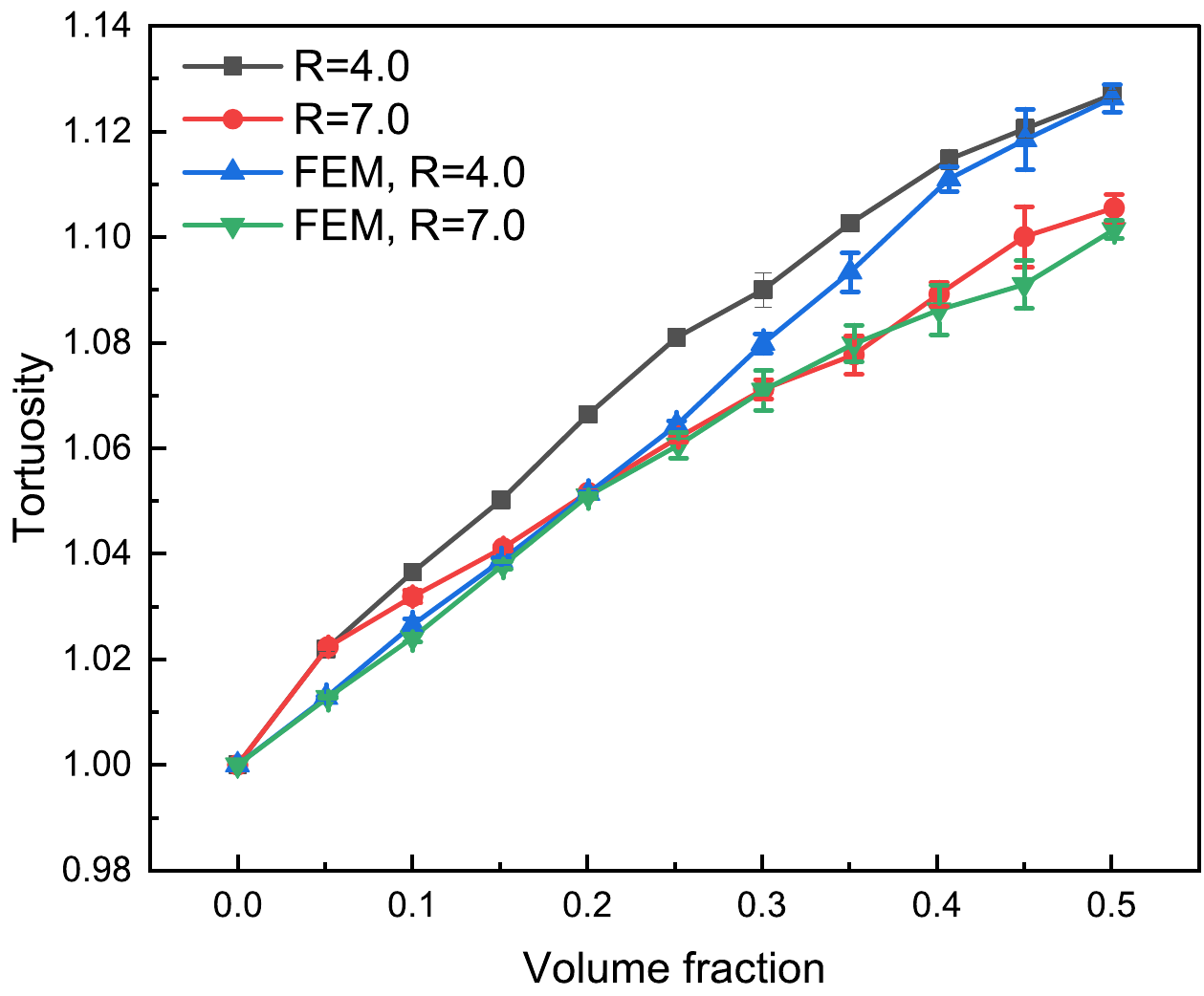}
   \caption{Relationship between porosity-particle size-tortuosity.}
   \label{fig:Fig12_porosity_size_tortuosity}
\end{figure}

The pores in a porous structure are directly influenced by the size of particle that make up the porous structure, and the value of tortuosity is closely related to the complexity of the pores. In order to investigate the effect of different particle sizes on the value of tortuosity, two different particle sizes are selected ($R=4, R=7$), and the results are shown in \cref{fig:Fig12_porosity_size_tortuosity}. Both FEM and TBM show that for structures with lower volume fractions of particles, particle sizes have almost no influence on the tortuosity. This can be explained by the fact that, in structures with high porosity, the pore phase occupies much of the space than particles, and the particle phase thus has limited contribution to the overall transport property of the structure. Therefore, for these structures, even though the size of particle (size of obstacles) is changed, the effect on the tortuosity is not significant because it is not playing a dominant role. However, in structures with a high volume fraction of particles, particle size does influence tortuosity: the larger the particle size, the lower the tortuosity. This is because structures with smaller particles are more likely to form narrower and more complex pore structures, leading to higher tortuosity of the structures. This agrees with Joseph \`Avila et al., which indicates that the greater the average pore size, the lower the tortuosity~\citep{Joseph_2022_SR23643}. In traditional empirical models (\cref{tb:empirical_models}), it is assume that tortuosity is only a function of porosity, disregarding the effect of changes in particle size or pore size on tortuosity. The results of this study demonstrate that, in addition to porosity, tortuosity is also correlated with the particle size (pore size) of the particles. 

\subsection{Effect of particle aggregation on tortuosity}
\begin{figure}[h]
   \centering
   \includegraphics[scale=0.45]{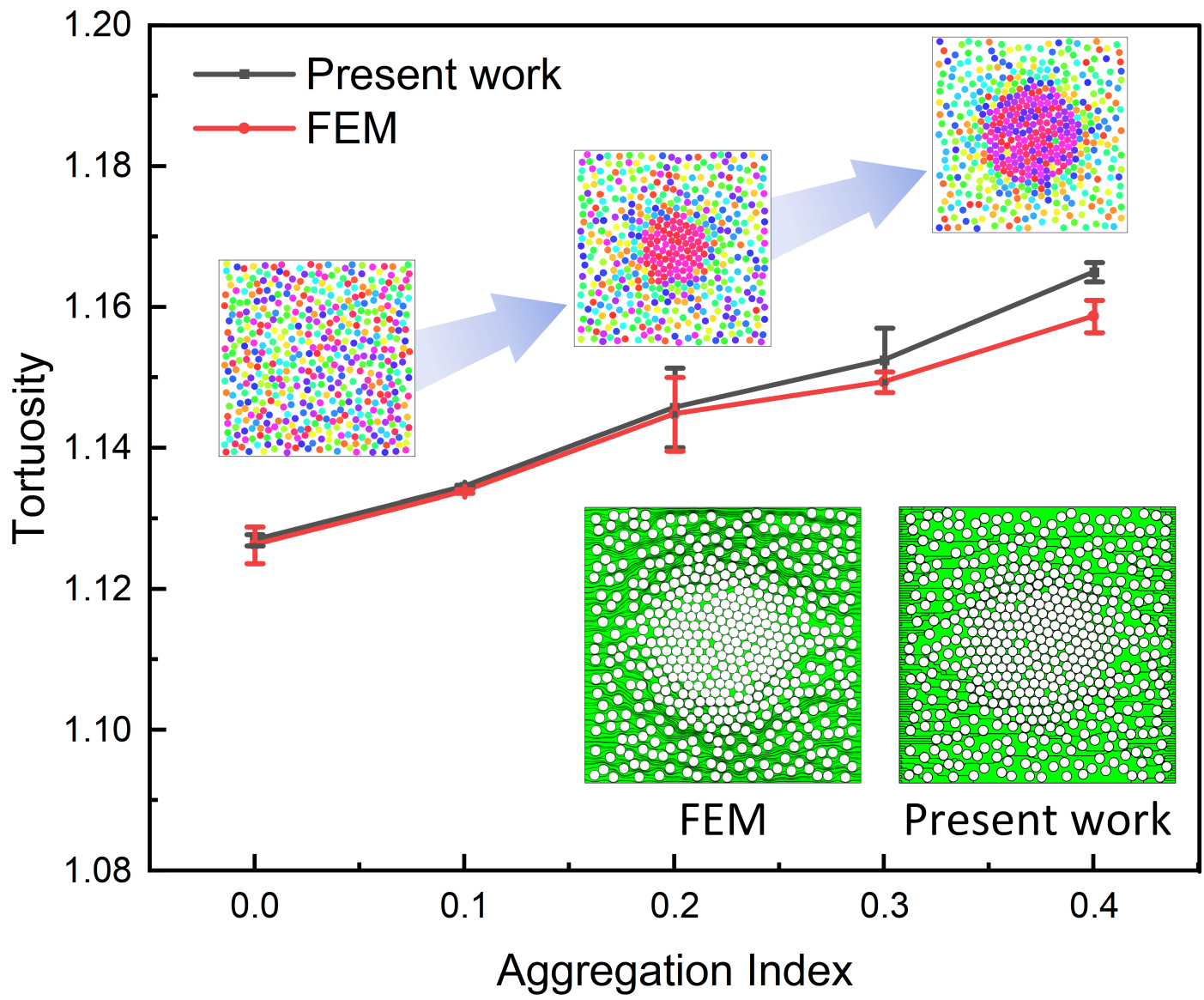}
   \caption{Relationship between the aggregation index and the tortuosity.}
   \label{fig:Fig13_Aggregation_index}
\end{figure}

In this section, the effect of solid-phase morphology in porous structures on tortuosity is further explored. In the aggregation, the particles are in contact with each other and leave no electrolyte in between, thus no ionic flux is allowed to go through. The aggregation index is characterized by the DBSCAN explained in \cref{sec:Aggregation morphology recognition}. For the calculation, the radius of the particles is set as $R=4$, and the volume fraction of the particles is $50\%$. Based on the definition of particle aggregation described above, porous structures with different particle aggregation morphologies are established, and their tortuosity is evaluated. \Cref{fig:Fig13_Aggregation_index} shows the results of tortuosity for porous structures with a single aggregated block of different degree. Both FEM method and our method show the same results, especially when the aggregation index is low. However, due to the narrow gaps between the aggregated particles, it requires excessively fine meshes to resolve the gaps geometrically, which would greatly increase the time required for mesh generation. Therefore, our method is much more suitable for a fast estimation of the tortuosity for aggregated particles. As for the value of tortuosity, it can be observed from the figure that as the aggregation index of the particles in the two-dimensional porous structure increases (the number of particles involved in aggregation increases), the tortuosity becomes higher. This is because larger aggregates increase the path length required for ions to bypass, thereby leading to an increase in tortuosity.

\begin{figure}[H]
   \centering
   \includegraphics[scale=0.45]{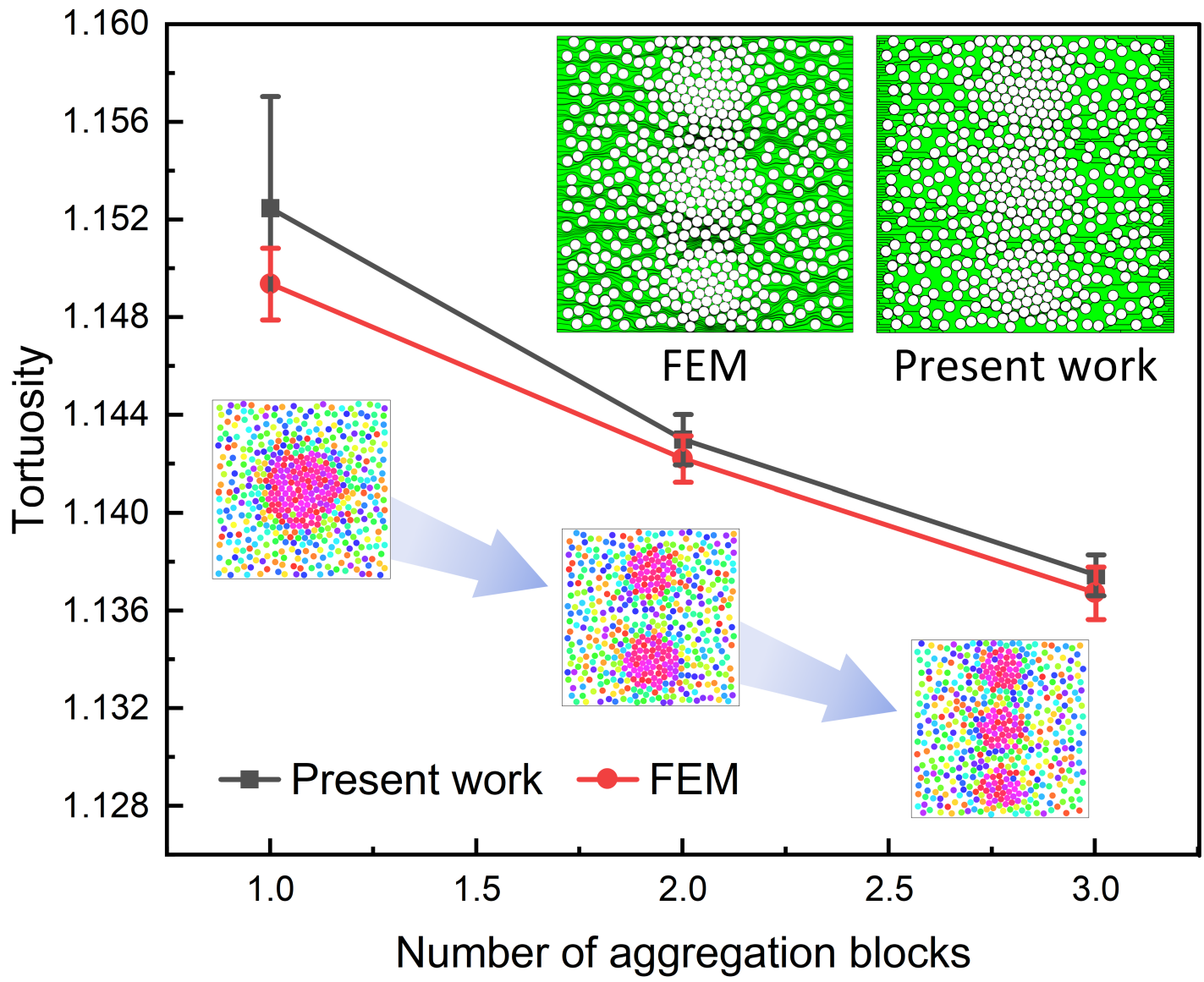}
   \caption{Relationship between the number of aggregation blocks and the tortuosity.}
   \label{fig:Fig14_aggregation_blocks}
\end{figure}

We then proceed to study the effect of variation in the number of aggregation blocks on the tortuosity. The aggregation index of the particles is fixed at $30\%$, and the number of aggregation blocks varies from 1 to 3. The tortuosity variation of the porous structure against different number of aggregation blocks is shown in \cref{fig:Fig14_aggregation_blocks}. It can be observed from the figure that as the number of aggregation blocks increases, i.e., the size of the aggregations decreases, the value of the tortuosity continuously decreases. The reason for this is that the larger the aggregation block, the longer the path required for the ions to bypass the aggregation fast, which in turn leads to greater tortuosity. In contrast, smaller but more numerous aggregation block significantly reduce the ionic diffusion paths compared to larger aggregation block, resulting in a smaller value of tortuosity. Combining \cref{fig:Fig12_porosity_size_tortuosity,fig:Fig13_Aggregation_index,fig:Fig14_aggregation_blocks}, it can be concluded that the tortuosity of porous structures is not only related to porosity and particle size, but is also closely related to the morphology of the solid phase in the porous structure. In the design process of porous structures (such as electrochemical devices), in order to obtain a structure with lower tortuosity, it is necessary to simultaneously consider the influence of the above variables on the value of tortuosity.

\subsection{Correction of P2D model based on tortuosity estimation}
The P2D model is currently a widely used model for simulating lithium-ion batteries, which smears out the microstructure and uses effective parameters at the cell level~\citep{Lamorgese_2018_JES011,Hashemzadeh_2024_JES104909}. Therefore, the effective conductivity $\sigma_l^\mathrm{eff}$ and diffusivity $D_l^\mathrm{eff}$ in the composite electrode have an important effect on the computational accuracy of the P2D model. Currently, the effective conductivity $\sigma_l^\mathrm{eff}$ and diffusivity $D_l^\mathrm{eff}$ in the composite electrode are related to the tortuosity $\tau$ and porosity $\varepsilon$ by the following equation~\citep{AN_2021_EA137775}:

\begin{align}
\sigma _l^\mathrm{eff} = {\sigma _l} \cdot \frac{{{\varepsilon _l}}}{\tau}
\label{eq:electron_effective}
\end{align}
\begin{align}
D_l^\mathrm{eff} = {D_l} \cdot \frac{{{\varepsilon _l}}}{\tau }
\label{eq:ion_effective}
\end{align}
where $\sigma_l$, $D_l$, $\varepsilon _l$ denote the conductivity, Li-ion diffusivity, and porosity of the pore phase, respectively.

\begin{figure}[H]
   \centering
   \includegraphics[scale=0.55]{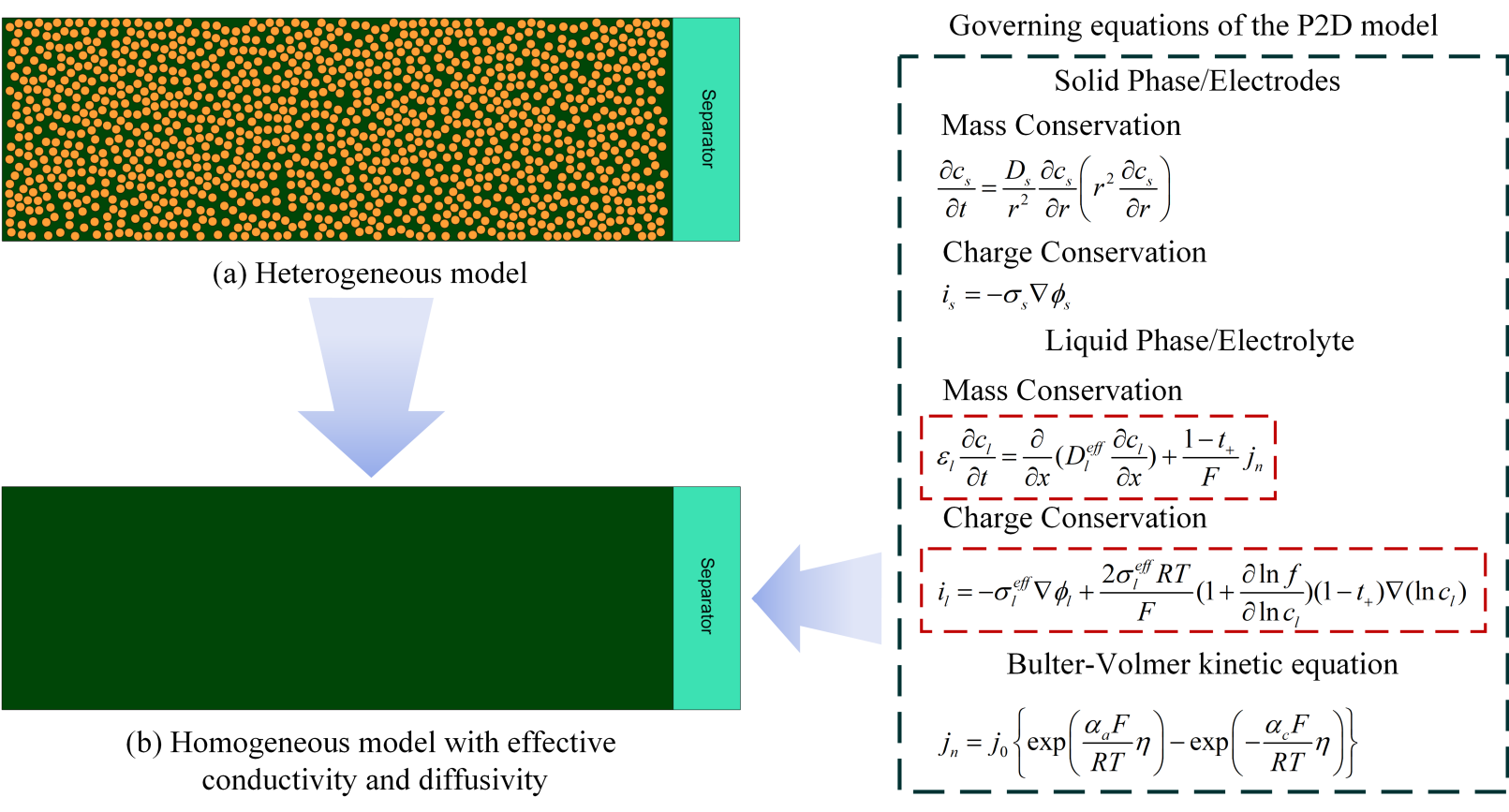}
   \caption{Schematic of the electrode models and the equation in the red box is used for coefficient correction.}
   \label{fig:Fig15_Schematic_of_electrode}
\end{figure}

Traditionally, the Bruggeman relation  (Eq.\ref{model:3}) is often used to estimate the tortuosity of porous electrodes. However, recent studies have shown inconsistent findings regarding the effectiveness of the Bruggeman relationship: while tomography-based microstructure simulations agree with the Bruggeman correlation in some cases, significant discrepancies exist in others~\citep{TJADEN_2016_COCE006,Usseglio_2020_JES913b}. In particular, when dealing with low porosity electrodes, the Bruggeman relation struggles to accurately reflect their complex tortuosity~\citep{Usseglio_2020_JES913b,Francois_2018_JES0731814}. Therefore, it is important to recognize the limitations of the Bruggeman relation in lithium battery simulation studies. In this section, the heterogeneous  model of porous electrodes and the corresponding P2D model are developed, as shown in \cref{fig:Fig15_Schematic_of_electrode}. The tortuosity of the heterostructured model is estimated by TBM, which is then substituted into \cref{eq:electron_effective,eq:ion_effective} for the calculation of effective electronic conductivity and ionic diffusivity, respectively. As input parameters to the P2D model, the corrected transfer coefficients of electrons and ions are used for the simulation of a half-cell. Specifically, the electrode has a length of \SI{450}{\micro\meter} and a width of \SI{150}{\micro\meter}. The size of the active particle is $R=\SI{3}{\micro\meter}$, and the electrode porosity is 50$\%$. Other parameters used for electrochemical simulation are derived from our previous work~\citep{CHEN_2024_JPS234095}.

\begin{figure}[H]
  \centering
  \subfigure[0.5C]{
    \centering
    \includegraphics[scale=0.30]{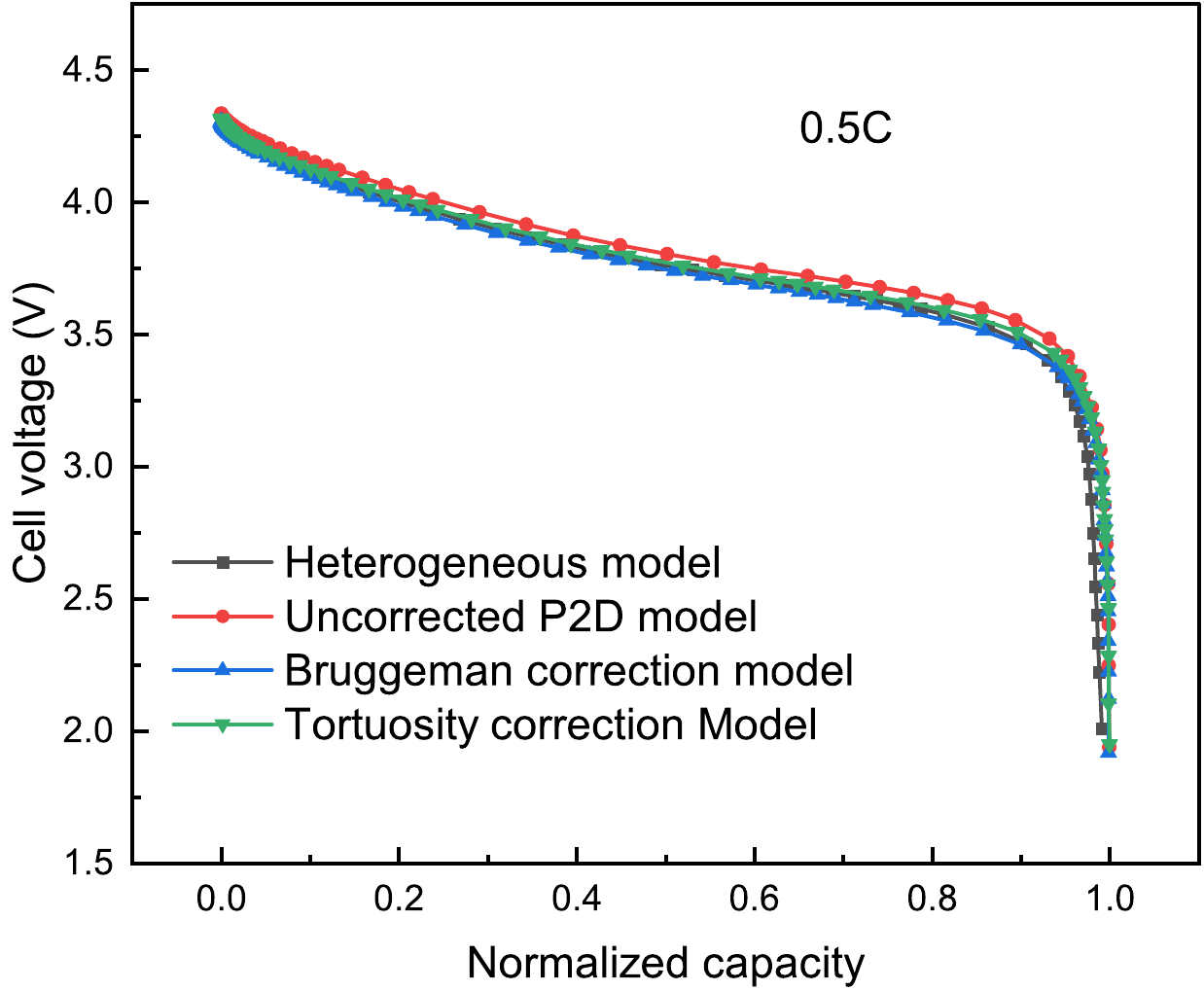}\label{fig:Fig16a_05C}
  }
\hspace{1.0cm}
    \subfigure[1C]{
    \centering
    \includegraphics[scale=0.30]{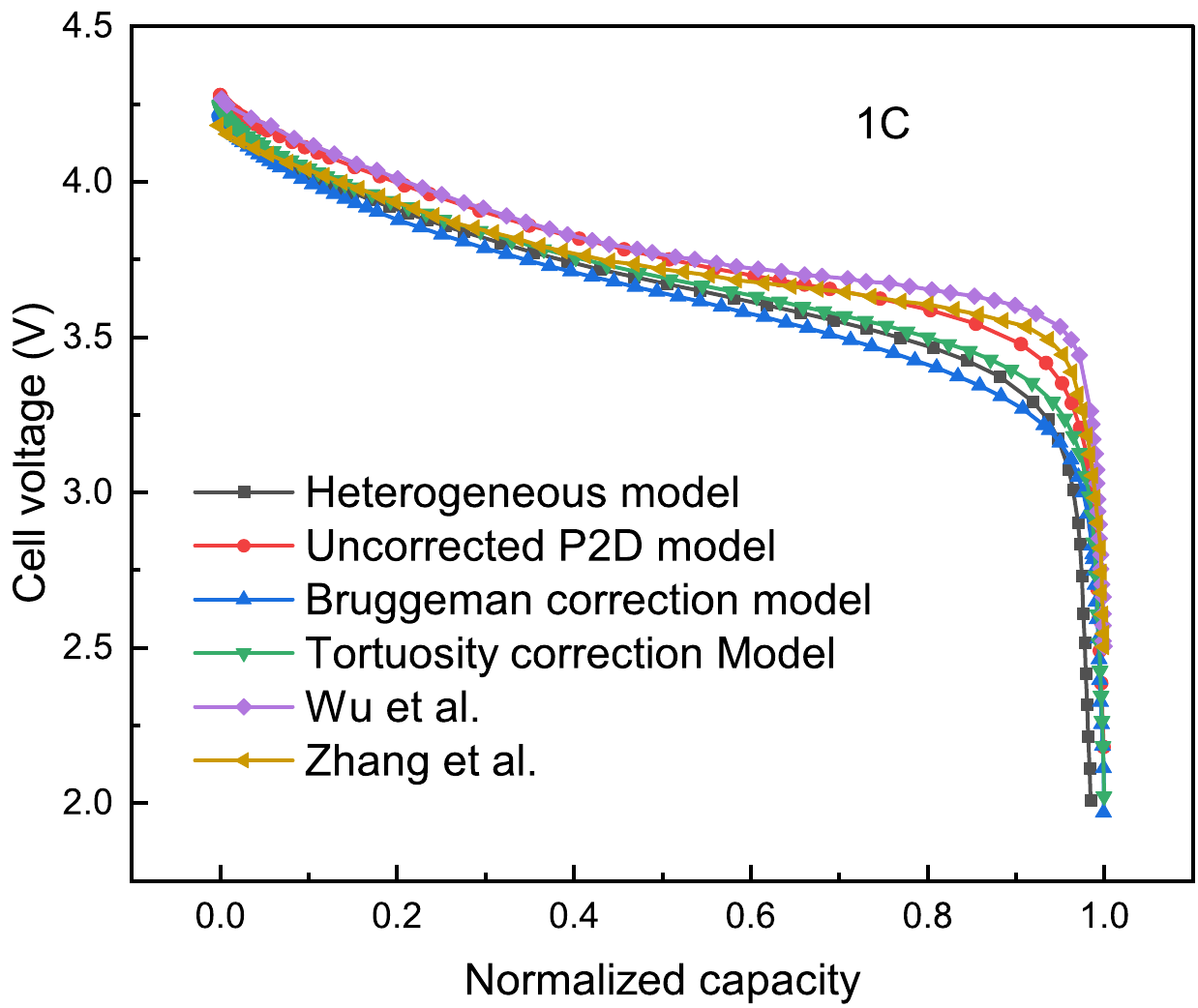}\label{fig:Fig16b_1C}
  }
  \subfigure[3C]{
    \centering
    \includegraphics[scale=0.30]{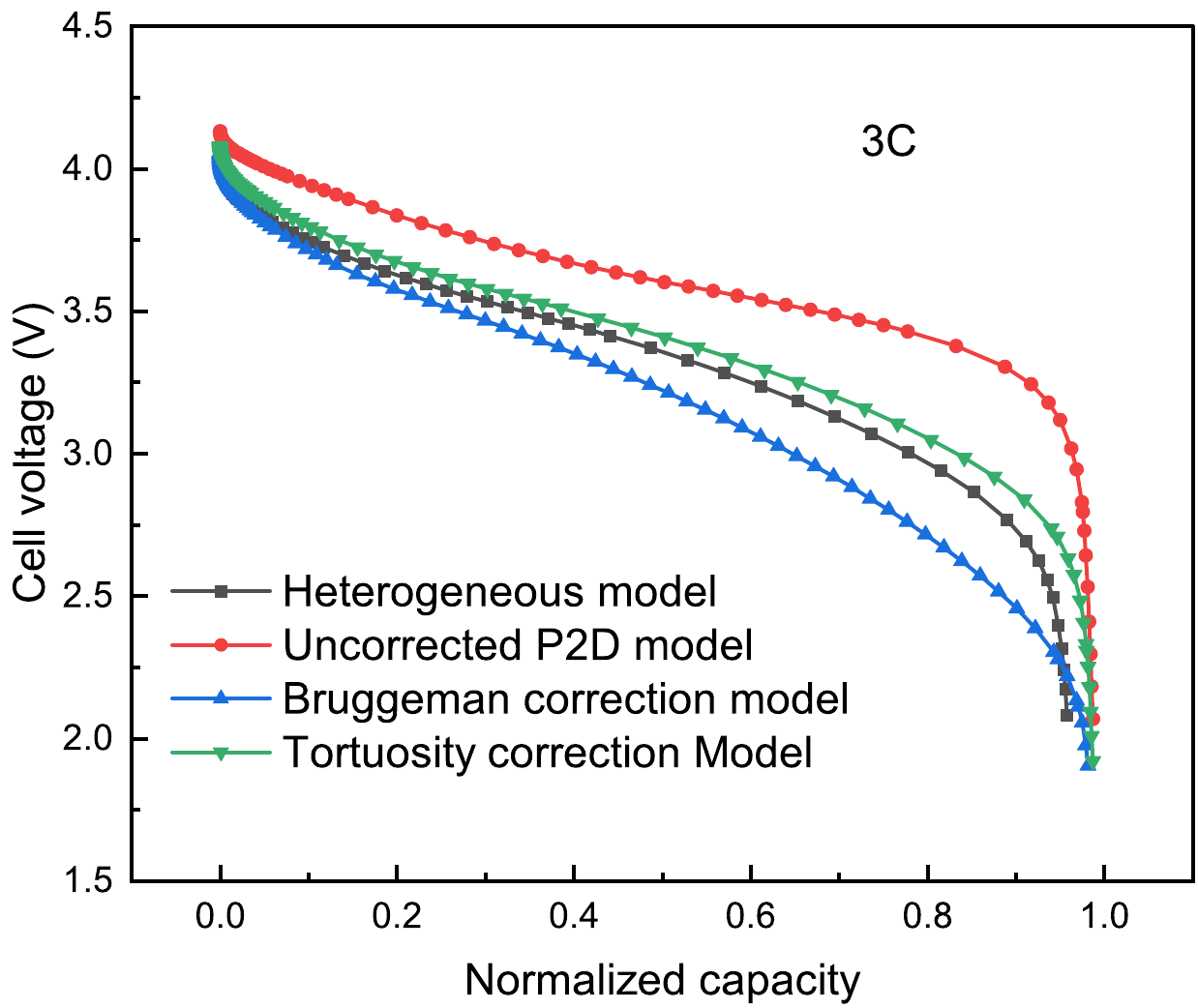}\label{fig:Fig16c_3C}
  }
\hspace{1.0cm}
    \subfigure[5C]{
    \centering
    \includegraphics[scale=0.30]{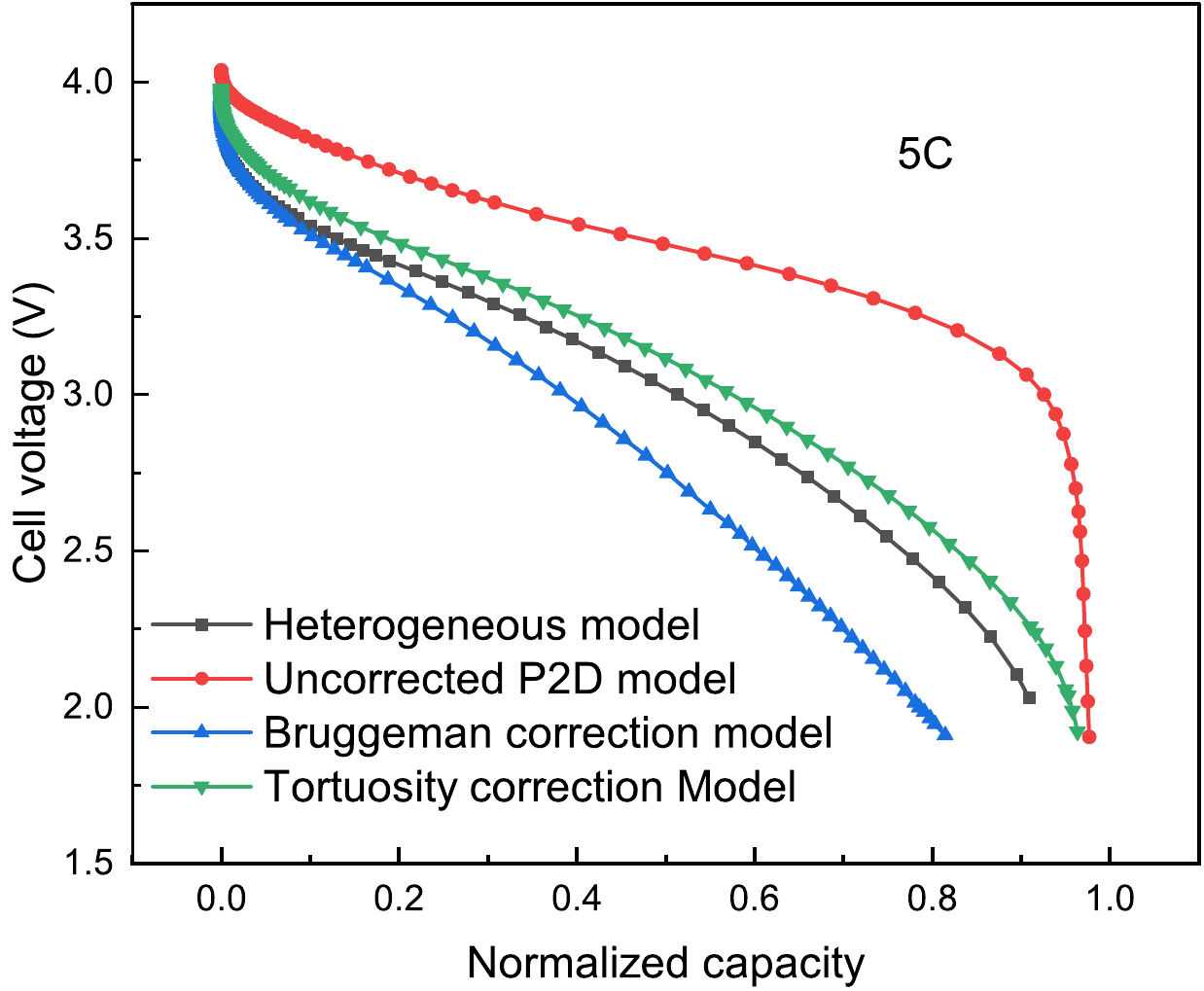}\label{fig:Fig16d_5C}
  }
 \caption{Variation of discharge curves at different discharge rates}
 \label{fig:Fig16_Variation_of_cell_voltage}
\end{figure}

\Cref{fig:Fig16_Variation_of_cell_voltage} shows the prediction of the discharge curves at different C-rate for a half cell with different. In the P2D models, we also use Bruggeman relation (Eq. (\ref{model:3})) and uncorrected tortuosity ($\tau = 1$) for calculation of effective coefficients for comparison. As a reference, we resolve the heterogeneous electrode as shown in \cref{fig:Fig15_Schematic_of_electrode}(a) with FEM. We compare the results of our simulations with the experimental results of Wu et al.~\citep{Wu_JES2012062204} and Zhang et al.~\citep{Zhang_ET20113654205}. It should be noted that the experimental cells are coin cells, and the materials used for the active particles and the electrolyte are the same as those used in our simulations, while the particle sizes of the active particles are different. Considering the consistency of the materials used, we believe the comparison is meaningful. Due to lack of experimental data from the same battery system, we only compared the discharge results at 1C discharge rate. As shown in \cref{fig:Fig16b_1C}, at a constant current discharge rate of 1C, the discharge curve obtained from our model closely matches the experimental results. The slight differences between the two are primarily due to minor inconsistencies in particle size. Nevertheless, this result fully validates that the heterogeneous model we adopted can accurately simulates the discharge characteristics of the battery. It can be observed that the P2D model has good simulation accuracy at low discharge C-rate even without the correction of effective transfer coefficients, as shown in \cref{fig:Fig16a_05C,fig:Fig16b_1C}. The reason is that at lower discharge current densities, the transfer of electrons and ions in the pore phase is relatively slow, and the tortuosity has limited impact on the transport of substances. The effect of tortuosity on material transport gradually increases with increasing discharge rate, and the P2D model without parameter correction will exhibit a large computational error, which is more pronounced at higher discharge rates, as shown in \cref{fig:Fig16c_3C,fig:Fig16d_5C}. It can also be observed that the accuracy of the P2D model using the traditional Bruggeman relation with parameter correction is somewhat improved, but the results are still hardly satisfactory. However, the P2D model using this paper's method to estimate the tortuosity and perform the transmission parameter correction exhibits higher accuracy, and the accuracy obtained by this paper's method is much better than that of the traditional Bruggeman relation's correction results even at really higher discharge rate, as shown in \cref{fig:Fig16d_5C}. In summary, combined with \cref{fig:Fig9_time_coms} and \cref{fig:Fig16_Variation_of_cell_voltage}, the method in this paper provides a fast and accurate result for tortuosity estimation and can be effectively applied to the correction of P2D models.

\section{Conclusions}
In this work, a new method for evaluating the tortuosity of porous structures is proposed. The model employs Radical tessellation of the topological partition of porous structures, which can conveniently obtain the connectivity paths of pores in porous structures composed of circular (2D) or spherical (3D) particles. By using radial tessellation to perform topological division on porous structures, all paths in the porous structure that one object can flow or diffuse through are obtained. On this basis, the shortest path through the porous structure is found by Dijkstra search algorithm and the tortuosity of the porous structure is thereby calculated.

In order to enhance the accuracy of tortuosity estimation with loosely packed porous structure, a group of pseudo background particles is introduced. The results show that adding background particles can effectively help to accurately estimate tortuosity. It is also indicates that compared to the finite element method, the tessellation-based method with background particles has higher computational efficiency. In addition, this work suggests that when the diameter of background particles is less than or equal to one-fourth of the particle size that makes up the porous structure, it can ensure the accuracy of tortuosity estimation as well as computational efficiency.

Applying the algorithm of this paper to estimate tortuosity in different heterogeneous porous structures, the results indicate that tortuosity is not only correlated with the volume fraction (porosity) but also with the particle size and particle aggregation morphology. Finally, different forms of P2D parameter correction models are compared, and the results shown that TBM is very effective in improving the prediction accuracy of P2D models.

It is worth mentioning that although the model proposed in this article is currently only suitable for estimating the tortuosity of porous structures composed of circular (spherical) particles. However, if the tessellation method is successfully applied to the topological divisioning of systems with non-circular (non-spherical) particle packing structures, the method has a promising application in the particles with various shapes.
\section*{Acknowledgements}
This work is supported by National Natural Science Foundation of China (Project No. 12102305), Fundamental Research Funds for the Central Universities and the Shanghai Pujiang Project (22PJ1413200). Finally, we would like to thank Ninglin Du from Advanced Intelligent Systems NIO Inc. for the help in this work.




\end{document}